\documentclass[prd,twocolumn,reprint,preprintnumbers,nofootinbib,superscriptaddress,longbibliography]{revtex4-1}

\usepackage{soul}
\usepackage{verbatim}
\usepackage{booktabs}
\usepackage{siunitx}
\usepackage{hyperref}
\usepackage{tabularx}
\usepackage{tikz}
\usepackage{subfigure}
\usetikzlibrary{tikzmark, calc}
\usepackage{amsmath}
\usepackage{amsthm}
\usepackage{amssymb}
\usepackage{graphicx}
\usepackage{ulem}
\makeatletter
\makeatletter \renewcommand{\@dotsep}{10000} \makeatother
\usepackage[utf8]{inputenc}
\usepackage{lmodern}
\usepackage{hyperref}
\usepackage[T1]{fontenc} 
\usepackage{amsmath}
\usepackage{graphicx, color}
\usepackage{physics}
\usepackage{url}

\usepackage{bm}
\usepackage{xcolor}
\usepackage{adjustbox}
\usepackage{tikz}
\usepackage{bbold}
\usetikzlibrary{quantikz2, positioning, calc}

\usepackage{amsmath,amssymb,bm,mathtools}
\usepackage{graphicx}
\usepackage{booktabs,tabularx,array}
\usepackage{xcolor}

\usepackage{enumitem}

\newcommand{\ii}{\mathrm{i}}
\newcommand{\vobs}{\vec  v_{\mathrm{obs}}}
\newcommand{\vrot}{\vec  v_{\mathrm{rot}}}
\newcommand{\vorb}{\vec  v_{\oplus}}
\newcommand{\rhat}{\hat{r}}
\newcommand{\Mbar}{\overline{M}}
\newcommand{\SHMpp}{\mathrm{SHM}^{++}}

\newcommand{\DKL}{D_{\mathrm{KL}}}
\newcommand{\Dil}{\mathcal D}

\begin{document}

\title{Dark Matter Velocimetry with Quantum Networks}

\author{Miguel A. Soto Alcaraz}
\email{msoto1@ku.edu}
\affiliation{Department of Physics and Astronomy, University of Kansas, Lawrence, KS 66045, USA}

\author{Doojin Kim}
\email{doojin.kim@usd.edu}
\affiliation{Department of Physics, University of South Dakota, Vermillion, SD 57069, USA}

\author{Kyoungchul Kong}
\email{kckong@ku.edu}
\affiliation{Department of Physics and Astronomy, University of Kansas, Lawrence, KS 66045, USA}
 
\author{Myeonghun Park}
\email{parc.seoultech@seoultech.ac.kr}
\affiliation{Center for Theoretical Physics of the Universe,
Institute for Basic Science, Daejeon 34126, South Korea}
\affiliation{Institute of Convergent Fundamental Studies, Seoultech, Seoul 01811, South Korea}
\affiliation{School of Natural Sciences, Seoultech,  Seoul 01811, South Korea}

\begin{abstract}
Quantum sensor networks can use the spatial phase coherence of wavelike dark matter to probe not only particle properties but also the motion and structure of the Galactic halo. 
We explore this idea for the minimal two-node network, in which a nonlocal exchange measurement samples the characteristic function of the velocity distribution. We compare the Standard Halo Model (SHM) with a Gaia-inspired radially anisotropic component and their mixture ($\SHMpp$), taking into account daily and annual modulation, postselection, depolarizing events, and a flat instrumental offset. 
We show that annual modulation at one fixed local sidereal
time has the same phase waveform for the centered SHM and
Gaia-inspired $\SHMpp$ models, leaving them degenerate when the overall signal amplitude is unknown. Earth’s rotation breaks this degeneracy by scanning the baseline through Galactic directions and exposing the direction-dependent coherence of an anisotropic population. 
Distinguishing these velocity distributions would test whether the local halo is smooth and nearly isotropic or contains a radially biased accreted component, turning a quantum-network detection into a probe of Milky Way structure. Our likelihood analysis shows how modulation and signal-odd controls separate this information from flat backgrounds and provides a path from model discrimination toward measurement of the local velocity distribution.
\end{abstract}

\maketitle


\section{Introduction}
\label{sec:introduction}

Cosmological observations require a dominant nonluminous matter component, but they do not specify its microscopic
identity~\cite{Planck2018}. The absence of a suitable particle candidate within the Standard Model (SM) makes dark matter (DM) a major motivation for searches for new physics. Despite decades of laboratory and astrophysical searches, its nature remains unknown. Proposed candidates span many orders of magnitude in mass
and interaction strength, ranging from ultralight particles to macroscopic objects such as primordial black
holes~\cite{Cooley:2022ufh,Bozorgnia:2024pwk}. This diversity motivates complementary detection strategies tailored to different candidates and their interactions.

Quantum sensing has opened new avenues within this broad search program. Recent proposals target dark photons~\cite{Chigusa:2025rqs,Chen:2022quj,He:2024sfi}, axions~\cite{Chen:2024aya}, other light DM candidates~\cite{Kim:2020coy,Ito:2023zhp,Linehan:2024btp,Chao:2024owf,Badurina:2025idj,Dong:2025mdk,Kim:2026mto}, and high-frequency gravitational waves \cite{Yi:2026iko}.  Quantum-control protocols have also been proposed to enhance
coherent signals and suppress selected noise sources~\cite{Chen:2023swh,Shu:2024nmc}.
Experimental advances include squeezed-state receivers and superconducting-qubit photon counters for axion
searches~\cite{HAYSTAC:2024jch,Braggio:2024xed}, as well as dark-photon searches using nonclassical cavity states, flux-tunable
cavities, and scalable qubit architectures~\cite{Agrawal:2023umy,Zhao:2025thg,Kang:2025kaf}. These efforts are supported by progress in transmon design, modeling, and
characterization~\cite{Moretti:2024xel,Labranca:2025bvc}. For a broader overview of quantum sensors in high-energy physics, see, e.g., Refs.~\cite{Chou:2023hcc,Moroi:2026ges}.

In this work, we focus on ultralight bosonic dark matter, including axions,
axion-like particles, and dark photons, whose large occupation
numbers permit a classical-wave description over macroscopic
distances~\cite{Arias2012,Hui2017,Irastorza2018}.
Many laboratory searches for these candidates are optimized
to establish excess power at a frequency fixed mainly by
the particle mass.
A detected signal would immediately motivate a second
measurement: how does the dark matter move through the
Galaxy and through the laboratory?

The local velocity distribution affects the coherence time and coherence length of a wave signal, the conversion between a spectral line and particle parameters, and the optimal spatial separation of phase-sensitive sensors.  
Beyond these implications for laboratory searches, the velocity distribution can carry imprints of the Milky Way's formation history, including past accretion and merger events~\cite{Necib:2018igl,Necib2019}. 
The conventional Standard Halo Model (SHM) is a smooth isotropic Maxwellian with an escape-speed cutoff.  Gaia observations instead reveal radially biased accreted stellar populations~\cite{Gaia2018,Helmi2018}.  Although the mapping from stars to dark matter is not unique, simulations and stellar data motivate a local anisotropic component~\cite{Necib:2018igl,Necib2019}.  The analytic $\SHMpp$ construction combines an SHM component with a radially anisotropic ``Gaia Sausage'' component \cite{Belokurov:2018vpl} and provides a useful controlled benchmark~\cite{Evans2019}.

The different velocity structures of these halo models can be probed through the joint annual--sidereal modulation of a phase-sensitive signal measured with an Earth-fixed baseline. The modulation provides two complementary advantages.  First, Earth's rotation and orbit imprint predictable sidereal
and annual harmonics on the dark-matter signal. These nonconstant components can be separated from a stationary additive readout offset, which contributes only to the constant mode.
Second, Earth’s rotation sweeps an Earth-fixed detector through different Galactic directions, so the modulation amplitude and waveform encode the dark-matter velocity distribution. Annual modulation has a long history in dark-matter searches~\cite{Drukier1986,Spergel1988,Freese2013} and directional axion and material-response studies have shown how daily information can probe the velocity distribution~\cite{Knirck2018,Heikinheimo2024}. Distributed quantum sensing provides a natural language
for combining phase-sensitive probes~\cite{Degen2017,Zhang2021}.
Experimental progress in quantum networking includes coherent state transfer, entanglement distribution,
and distributed quantum operations between spatially separated nodes across several physical platforms~\cite{Ritter2012,Kurpiers2018,Pompili2021,Knaut2024,Main2025}.
For wavelike dark matter, networks of phase-sensitive detectors can exploit spatial coherence to provide an interferometric observable that complements annual and sidereal modulation~\cite{Derevianko2018,Foster2021,Brady2022,Jiang2024}.

Ref.~\cite{Fukuda2025} showed that a nonlocal two-sensor measurement can retain the relative phase of a light-dark-matter excitation and probe the dark-matter wind.\footnote{Here, a nonlocal measurement refers to a joint two-sensor observable that probes the coherence between excitations at spatially separated nodes, rather than the excitation probability at either node individually. Such coherence can, in principle, be accessed through a coherent joint readout or reconstructed from appropriately correlated local measurements.}
We adopt this two-site observable to study the astrophysical information that it carries and its statistical interpretation.

The goals of this paper are:
\begin{enumerate}
 \item to derive the two-node observable and identify which Fourier component of the velocity distribution it measures, 
 \item to include the full annual--sidereal geometry of an Earth-fixed directed baseline, 
 \item to formulate a background-robust binary likelihood with depolarization, a flat offset, and an unknown effective response amplitude, 
 \item to quantify SHM rejection throughout the Gaia-inspired anisotropy parameter space, and
 \item to examine what additional scans are needed to turn discrimination into a measurement of the velocity distribution.
\end{enumerate}

This paper is organized as follows. 
Section~\ref{sec:observable} derives the two-node directional observable and relates it to the Fourier transform of the dark-matter velocity distribution. Section~\ref{sec:halo} introduces the SHM and Gaia-inspired velocity models and their transformation to the laboratory frame. Section~\ref{sec:modulation} examines annual and sidereal modulation, while Section~\ref{sec:statistics} performs the background-robust statistical analysis for discrimination of the SHM and the Gaia-inspired $\SHMpp$ distributions. Section ~\ref{sec:conclusions} is reserved for conclusions. Throughout this paper, we use natural units $\hbar=c=1$ except when velocities are quoted in $\mathrm{km\,s^{-1}}$.

\section{Directional Sensing with Two Nodes}
\label{sec:observable}

In this section, we develop the theoretical framework for probing the dark-matter velocity distribution with a two-node quantum sensor network. We begin by describing how a wavelike dark-matter field can induce excitations at two spatially separated sensors, with relative phase information that depends on the dark-matter velocity. We then show how this phase information can be obtained through measurements of the two-sensor system and establish its connection to the underlying velocity distribution. Finally, we discuss what velocity information can be extracted from these measurements and how it depends on the separation between the two nodes. 

\subsection{Wavelike dark matter and the two-node phase difference}
To probe the dark-matter velocity distribution with a two-node quantum sensor network, we first examine how the dark-matter field induces excitations at two spatially separated sensors. In particular, the relative phase between their excitation amplitudes---which we will show explicitly soon---depends on the dark-matter velocity and the separation between the nodes, providing the basis for directional sensing. To quantify this effect, we express the dark-matter field as a superposition of plane-wave components with different velocities and calculate their contributions to the sensor excitation amplitudes. 

Over one coherence domain, the non-relativistic dark-matter field can be written schematically as~\cite{Hui2017,Cheong2025} \begin{equation}
\begin{aligned}
 \phi(\vec  x,s)&=\operatorname{Re}\Phi_+(\vec  x,s),\\
 \Phi_+(\vec  x,s)&=\int\dd^3v\,\mathcal A(\vec  v)
 e^{-\ii\omega s+ \ii m\vec  v\cdot\vec  x},\\
 \omega(v) &=m+\frac12mv^2+\mathcal O(mv^4),
\end{aligned}
\label{eq:field}
\end{equation}
where $m$ is the dark-matter mass, $s$ is time within an interrogation, and $\mathcal A(\vec  v)$ is the complex amplitude density of the dark-matter field in velocity space. 
$\mathcal A(\vec  v)$ specifies the amplitude and random phase of the plane-wave component with laboratory-frame velocity $\vec  v$.
Its ensemble covariance, based on the assumption of statistically uncorrelated velocity modes, is determined by the local dark-matter density $\rho_\phi$ and the velocity distribution $f_{\rm lab}(\vec  v)$
\cite{Fukuda2025}
\begin{eqnarray}
 \big\langle\mathcal A(\vec  v)\mathcal A^*(\vec  v')\big\rangle
 &=&\frac{2\rho_\phi}{m^2} f_{\rm lab}(\vec  v)  \delta^{(3)}(\vec  v-\vec  v'), \label{eq:field-covariance}\\
 \int\dd^3v\,f_{\rm lab}(\vec  v)&=& 1, \label{eq:veclab}
\end{eqnarray}
where $\langle\cdots\rangle$ denotes an average over realizations of the stochastic dark-matter field at fixed laboratory conditions. 
The characteristic velocity scale $v_0$ sets frequency and wave-number spreads of order $mv_0^2$ and $mv_0$, respectively. 
For $v_0=238$~$\mathrm{km\,s^{-1}}$ and $m=1 \,\mu {\rm eV}$, the coherence time and length scales become
\begin{equation}
\begin{aligned}
\tau_c &\sim 1.04\ {\rm ms}\left(\frac{1 \mu{\rm eV}}{m}\right)\left(\frac{238 \,\,{\rm km \,s^{-1}}}{v_0}\right)^2, \\
\ell_c &\sim 249\ \mathrm{m}\left(\frac{1 \mu{\rm eV}}{m}\right)\left(\frac{238 \,\,{\rm km \, s^{-1}}}{v_0}\right).   \end{aligned}
\end{equation}
Such a macroscopic coherence length allows spatially separated sensors to probe the relative phase of the dark-matter field, motivating the two-node network considered below.

We envision two sensors at positions $\vec  x_1$ and $\vec  x_2$. Each has a ground state $|0\rangle$, an excited state $|1\rangle$, and transition angular frequency $\omega_0$, so the two-sensor system's free Hamiltonian is
\begin{equation} 
H_{0}=\omega_0\sum_{j=1}^2\sigma_j^+\sigma_j^-,
\end{equation}
where $\sigma_j^+=|1\rangle_j\langle0|_j$ and $\sigma_j^-=|0\rangle_j\langle 1|_j$ are the raising and lowering operators at node $j$. We henceforth refer to these effective two-level degrees of freedom as ``qubits'', without restricting their microscopic realization to a particular sensor technology. 
We model the weak, linear coupling of the dark-matter field to each qubit for the $|0\rangle\leftrightarrow|1\rangle$ transition by the following interaction-picture Hamiltonian, retaining only the near-resonant terms \cite{Fukuda2025}, 
\begin{equation}
 H_I(s)=\sum_{j=1}^2
 \left[
 g_jF_j(s)\Phi_+(\vec  x_j,s)
 e^{+\ii\omega_0s}\sigma_j^+
 +\mathrm{H.c.}
 \right].
 \label{eq:interaction}
\end{equation}
Here $F_j(s)$ is the time-domain interrogation envelope, taken to vanish outside $0\leq s\leq\tau$, and $g_j$ is an effective calibrated coupling that includes the qubit matrix element and the normalization convention for $\Phi_+$. The factor $e^{+\ii\omega_0s}$ arises from the interaction-picture evolution of the qubit raising operator. 
Combined with the time-dependent phase factor $e^{-i\omega s}$ in the dark-matter field expansion of Eq.~\eqref{eq:field}, it yields $e^{-i\Delta\omega s}$, where $\Delta\omega\equiv\omega-\omega_0$ is often referred to as the detuning. 
The Hamiltonian is intended as an effective description; its microscopic coupling depends on the dark-matter candidate and qubit technology.

We first set the two-qubit system $|\Psi\rangle$ to be in the ground state, i.e., $|\Psi(0)\rangle=\ket{00}$. To first order in this coupling, the state after an interrogation time $\tau$ is given by 
\begin{equation}
 |\Psi (\tau)\rangle
 =U_I(\tau,0)|00\rangle
 \simeq |00\rangle+b_1|10\rangle+b_2|01\rangle, 
 \label{eq:state}
\end{equation}
where $U_I$ is the usual time-ordered unitary evolution operator, 
\begin{equation}
    U_I(\tau, \tau_0)=\mathcal{T}\exp\left[ -\ii\int_{\tau_0}^\tau ds \, H_I(s)\right].
\end{equation}
Here $b_j=\int\dd^3v\,a_j(\vec  v)$ denotes the total excitation amplitude at node $j$ in a given field realization, 
where $a_j(\vec v)$ represents the contribution per unit velocity-space volume from the dark-matter field component with velocity $\vec v$. 
Expanding $U_I(\tau,0)$ to first order, we identify
\begin{equation}
 a_j(\vec  v)
 =-\ii g_j\mathcal A(\vec  v)
 e^{\ii m\vec  v\cdot\vec  x_j}
 \widetilde F_j(\Delta\omega),
 \label{eq:node-amplitude}
\end{equation}
where $\widetilde F_j$ defines the frequency response corresponding to the interrogation envelope $F_j(s)$:
\begin{equation}
 \widetilde F_j(\Delta\omega)
 \equiv\int_0^\tau\dd s\,
 F_j(s)e^{-\ii\Delta\omega s}.
 \label{eq:filter-function}
\end{equation}

It is often convenient to define the directed baseline, its unit direction, and the associated velocity-space Fourier vector as follows, to calculate the single-excitation probability:
\begin{equation}
 \begin{aligned}
  \Delta\vec  r&\equiv\vec  x_2-\vec  x_1,
  & \rhat&\equiv\frac{\Delta\vec  r}{|\Delta\vec r\,|},\\
  \vec  q&\equiv m\Delta\vec  r=q\rhat, \quad
  & q&\equiv|\vec  q \,|=m |\Delta\vec r\,|.
 \end{aligned}
 \label{eq:qdef}
\end{equation}
A plane-wave component with velocity $\vec v$ has spatial wave vector $\vec k=m\vec v$, giving rise to a phase difference $\vec k\cdot\Delta\vec r=\vec q\cdot\vec  v$ between the two nodes.
While $\vec k$ represents the dark-matter momentum, $\vec q$ is determined by the baseline and serves as the Fourier-conjugate variable to velocity.\footnote{Not to be confused with the momentum transfer $\vec q$ that is conventionally chosen in literature.}

For simplicity, we follow the approach taken in Ref.~\cite{Fukuda2025} and assume matched, calibrated node responses throughout the benchmark analysis, i.e., $g_1=g_2$ and $\widetilde F_1(\Delta\omega)=\widetilde F_2(\Delta\omega)$.
The common factors then cancel out for each velocity component, and thus we have
\begin{equation}
\frac{a_2(\vec  v)}{a_1(\vec  v)}
 =e^{\ii m\vec  v\cdot\Delta\vec  r}
 =e^{\ii\vec  q\cdot\vec  v} \, ,
 \label{eq:relative-phase}
\end{equation}
which fixes the relative phase of each velocity component. 
Thus Eq.~\eqref{eq:state} can be written as
\begin{equation}
 |\Psi(\tau)\rangle\simeq |00\rangle
 +\int\dd^3v\,a_1(\vec  v)
 \left(|10\rangle+e^{\ii\vec  q\cdot\vec  v}|01\rangle\right).
 \label{eq:weak-state}
\end{equation}
Using Eq.~\eqref{eq:field-covariance} and the matched-node assumption, the ensemble-averaged excitation probability is identical at both nodes:
\begin{equation}
 \big\langle|b_1|^2\big\rangle 
 =|g_1|^2\frac{2\rho_\phi}{m^2}
 \int\dd^3v\,f_{\rm lab}(\vec  v)
 |\widetilde F_1(\Delta\omega)|^2.
\end{equation}
As a concrete benchmark, we take matched, unit-height interrogation envelopes, $F_1(s)=F_2(s)=1$ for $0\leq s\leq\tau$. Substituting this shape into Eq.~\eqref{eq:filter-function}, we find
\begin{equation}
 \widetilde F_1(\Delta\omega)
 =\tau e^{-\ii\Delta\omega\tau/2}
 \frac{\sin(\Delta\omega\,\tau/2)}
      {\Delta\omega\,\tau/2}.
\end{equation}
When $|\Delta\omega \tau | \ll1$ over the velocity range where \(f_{\rm lab}(\vec v)\) is significant, the frequency response is approximately $\widetilde F_1(\Delta\omega)\simeq\tau$. 
Defining \(\epsilon^2\equiv |g_1|^2(2\rho_\phi/m^2)\), the ensemble-averaged excitation probabilities at the two nodes are identical, \(\langle|b_1|^2\rangle=\langle|b_2|^2\rangle\simeq\epsilon^2\tau^2\). The total probability of a single excitation induced by the dark-matter signal is then given by
\begin{equation}
 P_{\rm{1exc}}
 =\big\langle|b_1|^2+|b_2|^2\big\rangle 
 \simeq2\epsilon^2\tau^2.
\end{equation}
%

\subsection{Postselection and exchange measurements}

In the previous subsection, we obtained the ensemble-averaged probability $P_{\mathrm{1exc}}$ for a signal to produce one excitation. 
However, this quantity does not retain directional information about the dark-matter velocity distribution. Such information lies in the relative phase between the excitation amplitudes at the two nodes. To access this information, we condition on this outcome by projecting the two-qubit state in Eq.~\eqref{eq:state} onto the one-excitation
subspace with
\begin{equation}
    \Pi_1=|10\rangle\langle10|+|01\rangle\langle01|.
\end{equation}
This operation, called {\it postselection}, retains the coherence between $|10\rangle$ and $|01\rangle$ without distinguishing which node is excited. 
For an individual velocity component in Eq.~\eqref{eq:weak-state},
the matched-node phase relation gives the normalized state
\begin{equation}
 |\psi_{\vec  v}\rangle
 =\frac{|10\rangle+e^{\ii\vec  q\cdot\vec  v}|01\rangle}{\sqrt{2}}.
 \label{eq:poststate}
\end{equation}
While Eq.~\eqref{eq:poststate} describes the state associated with an individual velocity component, the dark-matter field consists of a stochastic superposition of many such components. To connect the underlying velocity distribution to measurable observables of the two-qubit system, we therefore construct the ensemble-averaged density matrix of the postselected state by averaging the contributions of the uncorrelated velocity components. As we will show, its off-diagonal elements encode the velocity-dependent relative phases and can be probed through exchange measurements.

For a given dark-matter field realization, projecting the two-qubit state onto the one-excitation subspace yields the {\it un}normalized state
\begin{equation}
 |\chi_{\mathcal A}\rangle
 \equiv \Pi_1|\Psi(\tau)\rangle
 =\sqrt{2}\int\dd^3v\,a_1(\vec  v)|\psi_{\vec  v}\rangle. \label{eq:chiAA}
\end{equation}
To obtain the density matrix describing the accepted signal events, we first average the unnormalized density operator $|\chi_{\mathcal A}\rangle\langle\chi_{\mathcal A}|$ over field realizations and then normalize the result. This procedure properly accounts for the realization-dependent probability of obtaining a single-excitation event. 
In the matched, broadband limit, where $g_1=g_2$ and $\widetilde F_1(\Delta\omega)\simeq\widetilde F_2(\Delta\omega)\simeq \tau$, Eq.~\eqref{eq:field-covariance} gives
\begin{equation}
 \big\langle a_1(\vec  v)a_1^*(\vec  v')\big\rangle
 \simeq \epsilon^2\tau^2
 f_{\rm lab}(\vec  v;d,h)\delta^{(3)}(\vec  v-\vec  v').
\end{equation}
Here we explicitly indicate the dependence of the laboratory-frame velocity distribution on the day of year $d$ and local
sidereal time (LST) $h$, measured in sidereal hours. The corresponding changes in the laboratory conditions are negligible over a single interrogation time. 
The covariance in the preceding equation implies that interference terms between distinct velocity components vanish upon ensemble averaging, leaving only contributions with \(\vec v=\vec v'\). We therefore obtain an unnormalized ensemble-averaged density matrix,
\begin{align}
 \big\langle
 |\chi_{\mathcal A}\rangle\langle\chi_{\mathcal A}|
 \big\rangle
 &=2\int\dd^3v\,\dd^3v'\,
 \big\langle a_1(\vec  v)a_1^*(\vec  v')\big\rangle
 |\psi_{\vec  v}\rangle\langle\psi_{\vec  v'}| \nonumber\\
 &\simeq 2\epsilon^2\tau^2
 \int\dd^3v\,f_{\rm lab}(\vec  v;d,h)
 |\psi_{\vec  v}\rangle\langle\psi_{\vec  v}|. \label{eq:unnormdensity}
\end{align}
We find that the trace of Eq.~\eqref{eq:unnormdensity} is essentially $P_{\mathrm{1exc}}\simeq2\epsilon^2\tau^2$ using Eq.~\eqref{eq:veclab} and $\Tr \ket{\psi_{\vec v}}\bra{\psi_{\vec v}}=1$.
Dividing $\langle \ket{\chi_{\mathcal{A}}} \bra{\chi_{\mathcal{A}}} \rangle$ by $P_{\rm 1exc}$ therefore yields the normalized conditional signal density
matrix. In the ordered basis $\{|10\rangle,|01\rangle\}$,
\begin{align}
 \widetilde\rho({\vec q};d,h)&=\int\dd^3v\,f_{\rm lab}(\vec  v;d,h)
 |\psi_{\vec  v}\rangle\langle\psi_{\vec  v}| \nonumber\\
 &=\frac12\int\dd^3v\,f_{\rm lab}(\vec  v;d,h)
 \begin{pmatrix}
  1&e^{-\ii\vec  q\cdot\vec  v}\\
  e^{+\ii\vec  q\cdot\vec  v}&1
 \end{pmatrix} \nonumber\\
 &=\frac12
 \begin{pmatrix}
  1&C_{\rm lab}^*(\vec  q;d,h)\\
  C_{\rm lab}(\vec  q;d,h)&1
 \end{pmatrix},
 \label{eq:density}
\end{align}
where the coherence is defined as
\begin{equation}
 C_{\rm lab}(\vec  q;d,h)
 \equiv\int\dd^3v\,f_{\rm lab}(\vec  v;d,h)
 e^{\ii\vec  q\cdot\vec  v}.
 \label{eq:characteristic}
\end{equation}

For a fixed baseline vector $\vec q$, the velocity dependence of the postselected two-qubit density matrix is entirely contained in the complex coherence $C_{\rm lab}(\vec q)$, which probes a single Fourier component of the dark-matter velocity distribution.
Since \(C_{\rm lab}\) is an average of complex phase factors \(e^{i\vec q\cdot\vec v}\), each with unit magnitude, its magnitude satisfies \(|C_{\rm lab}|\leq1\). The reduction in \(|C_{\rm lab}|\) arises from cancellations among the different velocity-dependent phases and quantifies the resulting loss of coherence between the two nodes. Meanwhile, the phase \(\arg C_{\rm lab}\) is sensitive to the mean dark-matter velocity projected along the baseline and, more generally, to asymmetries in the projected velocity distribution.
Importantly, the coherence depends only on the velocity component along the baseline, $v_\parallel\equiv\hat r\cdot\vec v$, because $\vec q\cdot\vec v=qv_\parallel$. Hence, even a velocity distribution with a large spread in transverse velocities can yield $|C_{\rm lab}|=1$, provided all velocity components contribute the same relative phase. In this case, the postselected state remains pure. 
The ensemble-averaged state is mixed whenever $|C_{\rm lab}|<1$ (i.e., $\Tr \tilde\rho^2 = \frac{1+|C_{\rm lab}|^2}{2} < 1$). 
When $|C_{\rm lab}|=0$, both eigenvalues are equal
($\lambda_\pm=\frac{1\pm|C_{\rm lab}|}{2}=\frac{1}{2}$), and the postselected state is maximally mixed within the one-excitation subspace.

To extract the velocity-dependent coherence $C_{\rm lab}$ from measurements of the two-qubit system, we introduce two observables that allow us to probe its real and imaginary parts. These are the exchange quadratures in the one-excitation subspace, defined as
\begin{align}
 M_X&=|01\rangle\langle10|+|10\rangle\langle01|,\\
 M_Y&=-\ii|01\rangle\langle10|+\ii|10\rangle\langle01|.
 \label{eq:quadratures}
\end{align}
Both operators have eigenvalues $\pm1$ in the accepted subspace, corresponding to the two possible outcomes of each measurement. Their expectation values, obtained by averaging over repeated measurements, provide direct access to the complex coherence $C_{\rm lab}$. 
For an exchange observable $M$,
the expectation value in the postselected signal state is
$\langle M\rangle=\operatorname{Tr}(\widetilde\rho M)$. In particular, we have 
\begin{equation}
 \langle M_X\rangle=\operatorname{Re}C_{\rm lab},
 \qquad
 \langle M_Y\rangle=-\operatorname{Im}C_{\rm lab}.
 \label{eq:quadrature-means}
\end{equation}
More generally, the measurement basis can be adjusted by introducing an analysis phase $\alpha$, allowing us to probe different combinations of the real and imaginary parts of $C_{\rm lab}$. The corresponding observable is 
\begin{eqnarray}
 M_\alpha &=& e^{-\ii\alpha}|01\rangle\langle10|
 +e^{\ii\alpha}|10\rangle\langle01| \\
 &=& \cos\alpha\,M_X+\sin\alpha\,M_Y.
 \label{eq:analysis-quadrature}
\end{eqnarray}
For each measurement of $M_\alpha$ in the postselected signal state, the outcome is $y=\pm1$. The probability of obtaining a given outcome, for a specified measurement phase $\alpha$, baseline $\vec q$, and observation time $(d,h)$, can be obtained by applying the Born rule with the corresponding operators $O_y=(I+yM_\alpha)/2$:
\begin{equation}
 \begin{aligned}
 P(y|\alpha,\vec q,d,h)
 &=\operatorname{Tr}(\widetilde\rho \,O_y)=\frac{1+y\langle M_\alpha\rangle}{2}\\
 &=\frac12\left\{1+y\operatorname{Re}
 \left[e^{\ii\alpha}C_{\rm lab}(\vec q;d,h)\right]\right\}.
 \end{aligned}
 \label{eq:quadrature-probability}
\end{equation}
The $\alpha=0$ and $\alpha=\pi/2$ choices measure $M_X$ and $M_Y$, respectively.  Operationally, $\alpha$ can be controlled through a local phase shift followed by balanced two-node interference and computational-basis readout.

An additional advantage of the adjustable analysis phase is that it allows us to reverse the sign of the measured signal without changing the detector baseline. 
For an Earth-fixed pair, one can  alternate between the analysis phases $\alpha=+\pi/2$ and $\alpha=-\pi/2$
in successive interrogations.
Their ideal signal expectations have opposite signs, so their difference removes an additive readout offset only if independent calibration establishes that the offset is the same in both settings. Relabeling previously recorded outcomes flips the offset along with the signal and provides no independent control.
Similarly, for a fixed pair of $(d,h)$, the relation $C_{\rm lab}(-\vec  q)=C_{\rm lab}^*(\vec  q)$ implies that the ideal $M_Y$ signal changes sign if the directed baseline is reversed.  

The analysis thus far has focused on the ideal coherent signal. To connect this formalism to realistic event-level forecasts, we introduce the incoherent accepted-event component in Section~\ref{sec:modulation}. Throughout the following analysis, all event counts refer to measurements accepted by projection onto the one-excitation subspace, unless stated otherwise. 

\subsection{Fourier-space interpretation}

Having established how the complex coherence $C_{\rm lab}$ can be measured, we now investigate what information about the dark-matter velocity distribution can be extracted from it. 
At fixed observation time $(d,h)$, $C_{\rm lab}$ is the characteristic function (or equivalently, Fourier transform) of the laboratory velocity distribution. It obeys
\begin{equation}
 C_{\rm lab}(\vec  0;d,h)=1,
 ~~~
 C_{\rm lab}(-\vec  q;d,h)=C_{\rm lab}^*(\vec  q;d,h).
 \label{eq:characteristic-properties}
\end{equation}
To understand its physical content, we first consider the short-baseline limit, where $|\vec q\cdot\vec v|\ll1$ over the relevant velocity range. 
Expanding $C_{\rm lab}$ in Eq.~\eqref{eq:characteristic} about $\vec  q=\vec  0$, we obtain various laboratory-frame velocity moments:
\begin{eqnarray}
 C_{\rm lab}(\vec  q;d,h)
 &=&1+\ii q_i\langle v_{i}\rangle_f
 -\frac12q_iq_j\langle v_{i}v_{j}\rangle_f \nonumber \\
&& -\frac{\ii}{6}q_iq_jq_k
   \langle v_{i}v_{j}v_{k}\rangle_f+\cdots,
 \label{eq:moment-expansion}
\end{eqnarray}
where $\langle g(\vec  v)\rangle_f\equiv
\int\dd^3v\,f_{\rm lab}(\vec  v;d,h)g(\vec  v)$ denotes the average over the laboratory-frame velocity distribution.\footnote{Here we used the Einstein summation convention for repeated indices \(i,j,k\), which run over three spatial components.} 
Using the relations in Eq.~\eqref{eq:quadrature-means}, we find
\begin{eqnarray}
 \langle M_Y\rangle
 &=&-q_i\langle v_{i}\rangle_f+\mathcal O(q^3), \\
 \langle M_X\rangle
 &=&1-\frac12q_iq_j\langle v_{i}v_{j}\rangle_f
  +\mathcal O(q^4).
\end{eqnarray}
At short baselines, $M_Y$ therefore probes the mean dark-matter velocity projected along the baseline, while the deviation of $M_X$ from unity probes the corresponding second moment. A finite baseline samples the characteristic function at its particular Fourier vector and can capture information beyond these leading moments. 
Measurements at multiple baseline lengths and orientations can therefore provide additional information needed to reconstruct the velocity distribution and help distinguish different Galactic halo models. 

To identify the baseline scale relevant for probing the dark-matter velocity distribution, we introduce the dimensionless Fourier radius
\begin{equation}
 x\equiv qv_0=mv_0 |\Delta\vec r\,|,
 \label{eq:xdef}
\end{equation}
where $v_0$ is the characteristic velocity scale of the distribution. With the $\ell_c=1/(mv_0)$ convention, this can also be expressed as $x=|\Delta\vec r\,|/\ell_c$.
For $x\ll1$, $C_{\rm lab}$ approaches unity and differences between velocity distributions appear only through small moment-dependent corrections. For a projected velocity spread of order $v_0$, the phases of different components largely cancel at $x\gg1$, suppressing $|C_{\rm lab}|$. Narrower velocity components, however, can remain coherent to larger $x$. These considerations suggest that $x\sim\mathcal O(1)$ is a useful design scale for choosing the baseline length, although the optimal radius depends on the velocity distributions, baseline direction, and the measurement noise.

\section{Velocity Models and the Laboratory Boost}
\label{sec:halo}

Having established the connection between the two-qubit coherence and the dark-matter velocity distribution, we now apply the formalism to concrete Galactic halo models. We consider the isotropic SHM and a Gaia-inspired model containing a radially anisotropic component, allowing us to investigate how different velocity structures affect the measurable coherence. Since the detector is located on Earth, we also need to incorporate the motion of the laboratory and the changing orientation of the baseline relative to the Galaxy into the associated calculation. Combining these ingredients, we derive the time-dependent response of an Earth-fixed two-node network, which provides the basis for the modulation analysis in Section~\ref{sec:modulation}. 

\subsection{Galactic convention, LSR, and SHM}

To describe the dark-matter velocity distribution in the Milky Way and relate it to the orientation of our detector, we introduce a Galactic coordinate system. We adopt a right-handed Cartesian basis in which $\hat{\bm X}$ points from the Sun toward the Galactic center, $\hat{\bm Y}$ points in the direction of Galactic rotation, and $\hat{\bm Z}$ points toward the north Galactic pole. This convention fixes the signs of the velocity components and the projections of the detector baseline onto the Galactic axes.

To determine the dark-matter velocity distribution observed by a terrestrial detector, we need to take into account the motion of the laboratory relative to the Galaxy. We begin with the Sun's motion, which can be decomposed into two components: the circular motion around the Galactic center and the Sun's additional motion relative to this circular orbit. 
The first component is described using the so-called local standard of rest (LSR), a reference frame moving on an ideal circular orbit at the Sun's distance from the Galactic center. In our Galactic coordinate system, its velocity is
\begin{equation}
    \vec v_{\rm LSR}=(0,v_c,0),
\label{eq:lsr}
\end{equation}
where $v_c$ is the circular speed, for which we adopt $v_c=238\ \mathrm{km\,s^{-1}}$~\cite{Baxter2021}. The Sun also has a small additional velocity relative to the LSR, known as its ``peculiar'' velocity, given by~\cite{Schoenrich2010}
\begin{equation}
\vec v_{\odot,{\rm pec}}
=(11.1,12.24,7.25)\ \mathrm{km\,s^{-1}}.
 \label{eq:solar-pec}
\end{equation}

Separately, we describe the dark-matter velocity distribution using the standard halo model (SHM), which assumes an isotropic Gaussian distribution in the Galactic rest frame, conventionally truncated at the escape speed $v_{\rm esc}$. Its width is characterized by the velocity parameter $v_0$, which is conceptually distinct from the circular speed $v_c$. For our numerical benchmark, we adopt $v_0=v_c=238\ \mathrm{km\,s^{-1}}$, following the conventional relation motivated by the ideal isothermal-sphere model~\cite{Evans2019}. The velocity-width parameter $v_0$ and circular speed $v_c$ describe different physical quantities and need not coincide in a more general Galactic model.

We neglect the escape-speed cutoff in our analytic benchmark. As we will see later, Fig.~\ref{fig:M_distributions} shows that it has only a small effect on the response curves
for the parameters considered. 
This approximation allows the velocity distribution and its Fourier transform to be expressed in particularly simple forms. The resulting SHM velocity distribution in the Galactic rest frame and its Fourier transform, also known as the velocity characteristic function, are
\begin{eqnarray}
 f_{\rm R}(\vec  u)
 &=&\frac{1}{\pi^{3/2}v_0^3}\,
 \exp\!\left(-\frac{u^2}{v_0^2}\right),\\
 \Phi_{\rm R}(\vec  q)
 &\equiv&\int\dd^3u\,f_{\rm R}(\vec  u)\, 
 e^{\ii\vec  q\cdot\vec  u} 
 =\exp\!\left(-\frac{q^2v_0^2}{4}\right).
 \label{eq:shm}
\end{eqnarray}
One should note that both the Galactic-frame velocity $\vec u$ and the baseline Fourier vector $\vec q$ in $\Phi_{\rm R}$ are expressed in the Galactic coordinate system. Since the detector baseline is fixed to the rotating Earth, its orientation in this coordinate system changes with observation time. This time dependence must be taken into account when relating $\Phi_{\rm R}$ to the laboratory-frame characteristic function $C_{\rm lab}$. The spherical escape-speed cutoff, $u<v_{\rm esc}$, and the corresponding normalization are discussed in Appendix~\ref{app:cutoff}.

\subsection{Gaia-Sausage component and \texorpdfstring{$\SHMpp$}{SHM++}}

To investigate whether the proposed quantum network can distinguish different Galactic dark-matter velocity distributions, we consider a second benchmark featuring velocity anisotropy, in contrast to the isotropic SHM. Specifically, we adopt an untruncated, Gaia-Sausage-inspired version of the $\SHMpp$ model in Ref.~\cite{Evans2019}, which incorporates a radially anisotropic component. The velocity distribution of this component, centered at zero velocity in the Galactic rest frame, is 
\begin{equation}
 f_{\rm S}(\vec  u)=
 \frac{\exp[-u_X^2/(2\sigma_r^2)
            -(u_Y^2+u_Z^2)/(2\sigma_t^2)]}
 {(2\pi)^{3/2}\sigma_r\sigma_t^2},
 \label{eq:sausage}
\end{equation}
where $\sigma_r$ is the one-dimensional dispersion along the local Galactic radial axis, while $\sigma_t$ is the dispersion along each of the two tangential axes. 
We quantify the degree of velocity anisotropy by
\begin{equation}
 \beta=1-\frac{\sigma_t^2}{\sigma_r^2}.
 \label{eq:beta}
\end{equation}
The isotropic limit corresponds to $\beta=0$, while positive values of $\beta$ indicate a broader velocity distribution in the radial direction than in the tangential directions. 

To isolate the effects of velocity anisotropy when comparing the two benchmarks, we require the Gaia-Sausage component to have the same total velocity dispersion as the untruncated SHM. Since both distributions are centered at zero velocity, this motivates us to match their second moments, i.e., $\langle u^2 \rangle $, 
\begin{equation}
    \langle u^2\rangle_{\rm S}
    =\sigma_r^2+2\sigma_t^2
    =\frac{3v_0^2}{2}
    =\langle u^2\rangle_{\rm R}.
\end{equation}
Combining this condition with Eq.~\eqref{eq:beta}, we obtain $v_0^2$-normalized (dimensionless) radial and tangential variances $a_r$ and $a_t$, 
\begin{equation}
 a_r\equiv\frac{\sigma_r^2}{v_0^2}
 =\frac{3}{2(3-2\beta)},
 \qquad
 a_t\equiv\frac{\sigma_t^2}{v_0^2}
 =\frac{3(1-\beta)}{2(3-2\beta)}.
 \label{eq:variances}
\end{equation}
In the isotropic limit, i.e., $\beta=0$, both dispersions $\sigma_r$ and $\sigma_t$ are reduced to $v_0/\sqrt{2}$, recovering $f_{\rm S}=f_{\rm R}$. The corresponding velocity characteristic function is
\begin{eqnarray}
 \Phi_{\rm S}(\vec  q)
 &\equiv&\int\dd^3u\,f_{\rm S}(\vec  u)e^{\ii\vec  q\cdot\vec  u} \nonumber \\ 
 &=&\exp\!\left[-\frac{v_0^2}{2}
 (a_rq_X^2+a_tq_Y^2+a_tq_Z^2)\right].
 \label{eq:sausage-characteristic}
\end{eqnarray}

We now construct our second benchmark, the $\SHMpp$ model, by combining the isotropic SHM distribution with the radially anisotropic Sausage component. 
Denoting the fraction of the local dark-matter density contributed by the Sausage component as $\eta$, the resulting velocity distribution and its characteristic function are given by
\begin{align}
 f_{++}(\vec  u)&=(1-\eta)f_{\rm R}(\vec  u)+\eta \, f_{\rm S}(\vec  u),\\
 \Phi_{++}(\vec  q)&=(1-\eta)\Phi_{\rm R}(\vec  q)
                   +\eta\, \Phi_{\rm S}(\vec  q).
 \label{eq:mixture}
\end{align}
The untruncated SHM is recovered in either the limit $\eta=0$, where the Sausage component is absent, or the limit $\beta=0$, where the Sausage component becomes isotropic and identical to the SHM distribution. 

To compare how the different halo models affect the two-qubit coherence, we express their characteristic functions in terms of the dimensionless baseline length in Eq.~\eqref{eq:xdef} and its direction $\hat r$ in the Galactic frame. 
Writing $\vec  q=(x/v_0)\rhat$, we define the corresponding coherence envelopes $A_i(x,\rhat)\equiv\Phi_i(\frac{x}{v_0}\rhat)$:
\begin{align}
 A_{\rm R}(x)&=e^{-x^2/4}, \label{eq:AR}\\
 A_{\rm S}(x,\rhat)&=
 \exp\!\left[-\frac{x^2}{2}
 \left(a_r\hat r_X^2+a_t\hat r_Y^2+a_t\hat r_Z^2\right)\right], \label{eq:AS} \\
 A_{++}(x,\rhat)&=(1-\eta)A_{\rm R}(x)+\eta A_{\rm S}(x,\rhat).
 \label{eq:model-envelopes}
\end{align}
The components of $\rhat$ here are expressed in the Galactic basis;
for an Earth-fixed baseline, they change with local sidereal time. 
We use $(\eta,\beta)=(0.3,0.9)$ as an illustrative Gaia-inspired benchmark. The Sausage fraction lies at the upper end
of the range $\eta=0.2\pm0.1$ considered in Ref.~\cite{Evans2019}, while the anisotropy
lies within its quoted $\beta=0.90\pm0.05$ range. With $v_0=238\ \mathrm{km\,s^{-1}}$, Eq.~\eqref{eq:variances} then gives
$\sigma_r=266.1\ \mathrm{km\,s^{-1}}$ and
$\sigma_t=84.1\ \mathrm{km\,s^{-1}}$.

\subsection{Observer velocity and the Earth-fixed response}

As introduced in Section~\ref{sec:observable}, we label the observation time by the day of year $d$ and LST $h$, measured in sidereal hours. In this subsection, all velocity vectors and baseline directions are expressed in the Galactic Cartesian basis.

The velocity of the apparatus relative to the Galactic frame is
\begin{equation}
 \vobs(d,h)=\vec  v_{\rm LSR}+\vec  v_{\odot,{\rm pec}}
 +\vorb(d)+\vrot(h),
 \label{eq:vobs}
\end{equation}
where $\vorb(d)$ is Earth's orbital velocity around the Sun and
$\vrot(h)$ is the velocity of the laboratory due to Earth's rotation.
We evaluate $\vrot$ at the baseline midpoint and neglect the difference
between the nodes' rotational velocities. The day $d$ specifies the
orbital phase; for a given site and baseline, LST specifies the
apparatus orientation relative to Galactic directions. Explicit expressions for Earth's orbital and rotational velocities, together with the transformation of the Earth-fixed baseline direction into Galactic coordinates, are given in Appendix~\ref{app:geometry}.

A dark-matter component with Galactic-frame velocity $\vec  u$ has
velocity $\vec  v=\vec  u-\vobs$ relative to the apparatus. Consequently, we have
\begin{equation}
 f_{\rm lab}(\vec  v;d,h)
 =f_{\rm gal}\!\left[\vec  v+\vobs(d,h)\right].
 \label{eq:lab-shift}
\end{equation}
Since both the SHM and $\SHMpp$ distributions are centered at zero in the Galactic frame, i.e., $\langle\vec u\rangle_{\rm gal}=0$, their mean velocities (i.e., first moments) in the laboratory frame are identical: 
\begin{equation}
 \langle\vec  v\rangle_f
 =-\vobs(d,h). 
 \label{eq:lab-mean}
\end{equation}
Here $\langle\cdots\rangle_f$ and $\langle\cdots\rangle_{\rm gal}$ denote averages with respect to $f_{\rm lab}$ and $f_{\rm gal}$, respectively. 

Changing variables from $\vec  v$ to $\vec  u$ in Eq.~\eqref{eq:characteristic}, we can express the laboratory-frame coherence $C_{\rm lab}$ for a given halo hypothesis $\mathcal H$ as
\begin{equation}
 C_{\rm lab}(\vec  q;d,h)
 =e^{-\ii\vec  q\cdot\vobs(d,h)}\Phi_{\mathcal H}(\vec  q).
 \label{eq:shift}
\end{equation}
Since the Galactic velocity distributions considered here are even functions of velocity, their characteristic functions $\Phi_{\mathcal H}$ are real. 
The two exchange quadratures in Eq.~\eqref{eq:quadrature-means} therefore have expectation values
\begin{eqnarray}
\langle M_X\rangle_{\mathcal H}
&=&\Phi_{\mathcal H}(\vec q)
\cos[\vec q\cdot\vobs(d,h)],\\
\langle M_Y\rangle_{\mathcal H}
&=&\Phi_{\mathcal H}(\vec q)
\sin[\vec q\cdot\vobs(d,h)].
\end{eqnarray}
Both observables carry information about the Galactic velocity distribution through $\Phi_{\mathcal{H}}$. In the following analysis, we focus on the $M_Y$ quadrature, which probes the projected dark-matter wind at leading order in the baseline separation. 
Hence, we define its ideal signal expectation value as
\begin{equation}
 \Mbar_{\mathcal H}(\vec  q;d,h)
 \equiv \langle M_Y\rangle_{\mathcal H} = \Phi_{\mathcal H}(\vec q)
\sin[\vec q\cdot\vobs(d,h)].
 \label{eq:response}
\end{equation}
For an Earth-fixed baseline, $|\Delta\vec r\,|$ and $x=mv_0|\Delta\vec r\,|$ are constant, while its direction in Galactic coordinates varies with local sidereal time: $\vec  q(h)=m |\Delta\vec r\,| \rhat(h)=(x/v_0)\rhat(h)$.
Substituting this relation into Eq.~\eqref{eq:response} and using the coherence envelopes defined in Eqs.~\eqref{eq:AR} through \eqref{eq:model-envelopes}, we obtain
\begin{equation}
 \Mbar_{\mathcal H}(x;d,h)
 =A_{\mathcal H}\bigl(x,\rhat(h)\bigr)
 \sin\!\left[
  \frac{x}{v_0}\vobs(d,h)\cdot\rhat(h)
 \right].
 \label{eq:response-x}
\end{equation}
The phase encoded in the sine factor measures the observer velocity projected along the baseline, while the envelope describes spatial dephasing associated with the Galactic velocity distribution. 
As Earth orbits the Sun, the changing observer velocity produces annual modulation of the signal. In the meantime, Earth's rotation continuously changes the baseline direction relative to the Galactic wind, leading to sidereal modulation through the projected velocity, and in the case of an anisotropic halo, through the coherence envelope itself.

\section{Annual and Sidereal Modulation}
\label{sec:modulation}

The key question is whether the annual and sidereal modulations mentioned above can distinguish different Galactic velocity distributions. For the even Galactic velocity distributions considered here, the SHM and $\SHMpp$ have the same projected-velocity factor in Eq.~\eqref{eq:response-x}, while the coherence amplitude $A_{\mathcal H}$ depends on the velocity distribution. We therefore examine how baseline choices and Earth's motion enable us to distinguish these two halo models.

\begin{figure*}[t]
    \begin{center}
    \vspace*{-0.6cm}
    \subfigure[]{\includegraphics[width=0.49\linewidth]{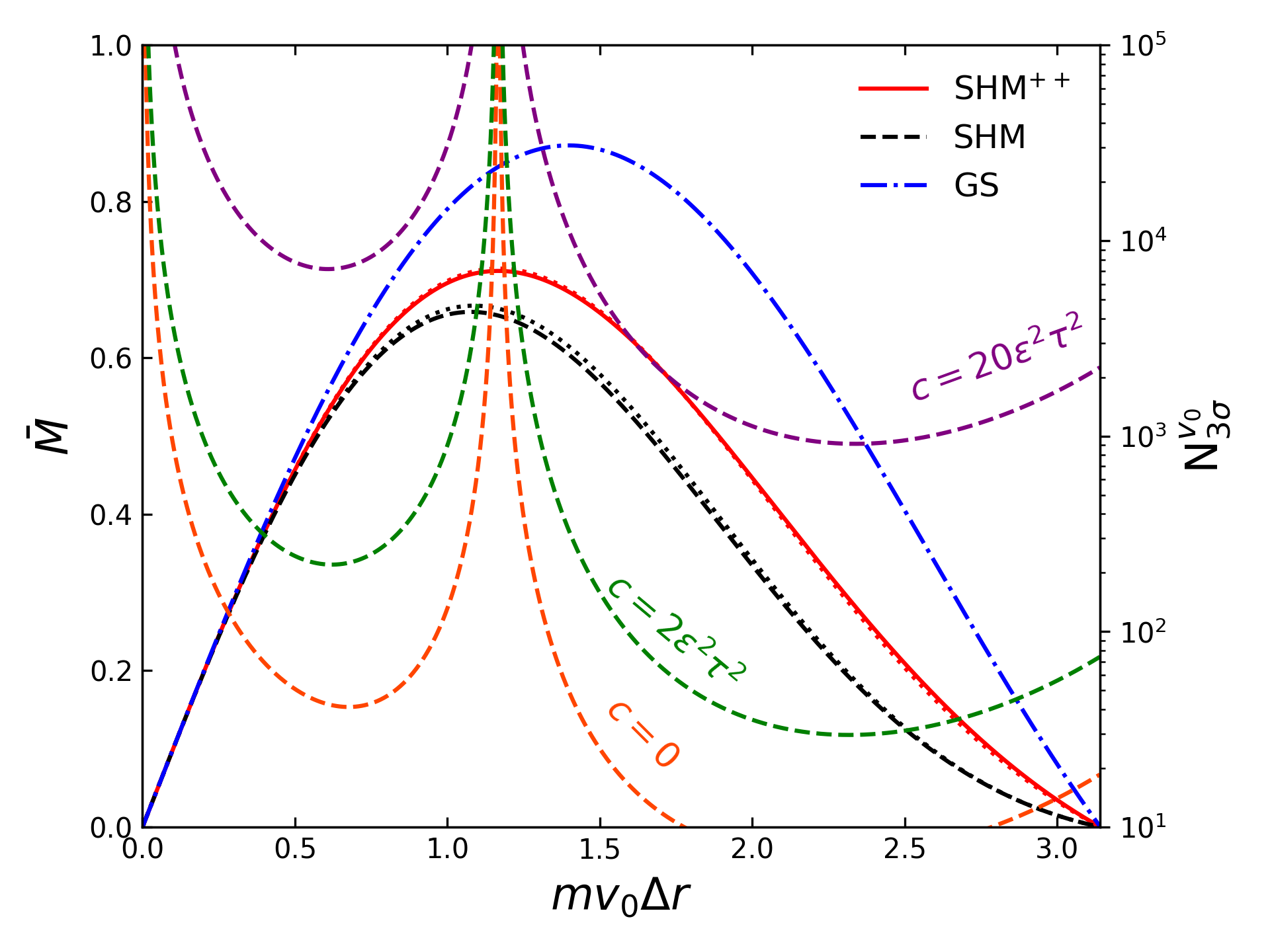}}
    \subfigure[]{\includegraphics[width=0.49\linewidth]{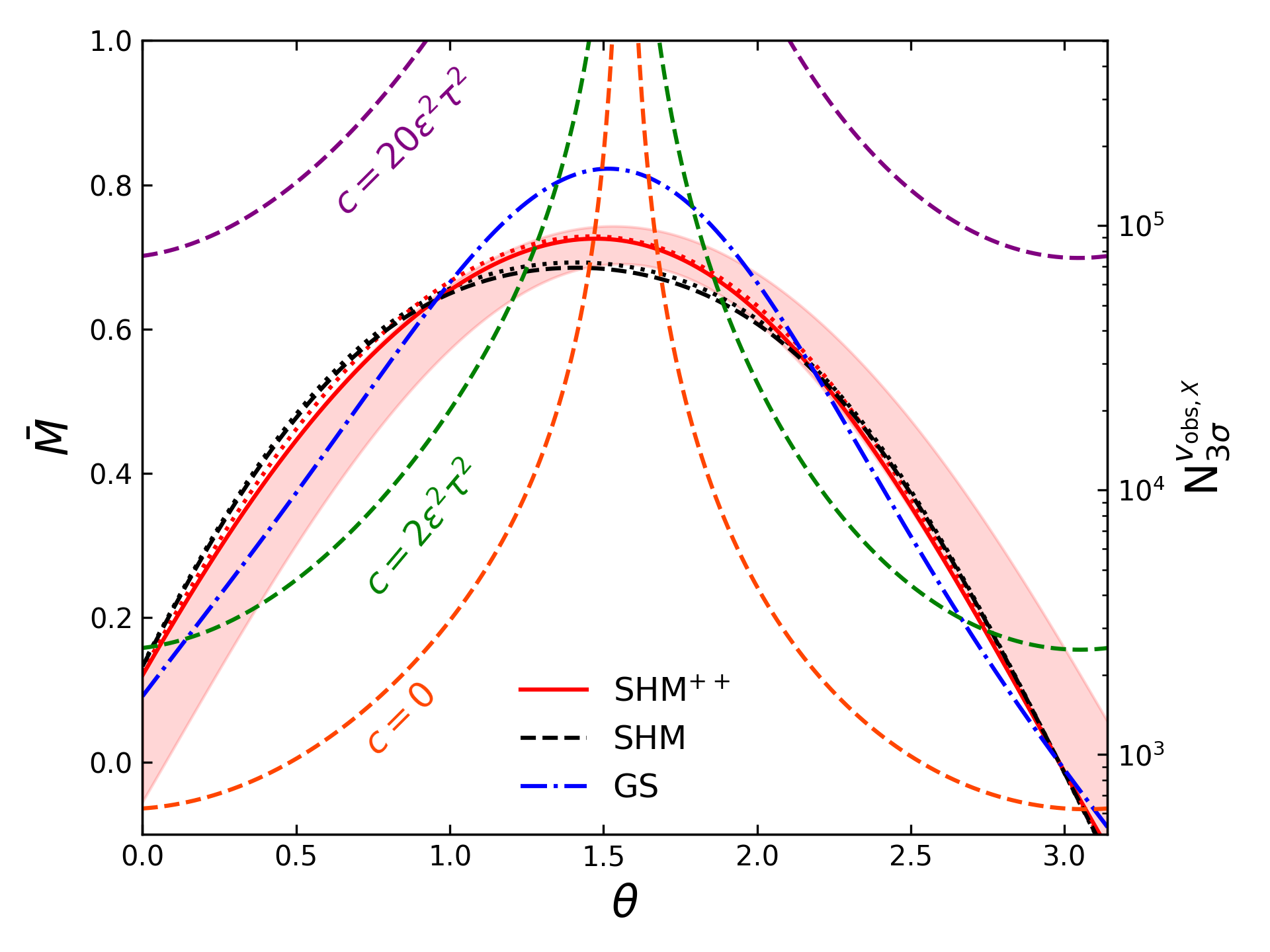}}
    \end{center}
    \vspace*{-0.5cm}
    \caption{$\Mbar$ response (on the left $y$-axes) and the required number of measurements (on the right-$y$ axes) for the different velocity distributions. In both panels, we use $\beta = 0.9$ for the Gaia Sausage and $\SHMpp$, and $\eta = 0.3$ for $\SHMpp$. The baseline detector is assumed to remain fixed in the Galactic frame, without accounting for Earth's rotation. Only the black and red dotted curves include the escape-speed cutoff $v_{\rm esc}=544\ \mathrm{km\,s^{-1}}$, corresponding to SHM and $\SHMpp$, respectively.    
    (a) The plots in the left panel show the expectation values $\Mbar$ as a function of $x=mv_0|\Delta\vec r\,|$ with the directed baseline along Galactic $+\hat{\bm Y}$ and $\vobs=v_0\hat{\bm Y}$. The solid red, dash-dotted blue, and dashed black curves represent $\SHMpp$, Gaia Sausage, and SHM, respectively. The right axis shows the number of measurements $N_{3\sigma}^{(v_0)}$ for $\SHMpp$ required to achieve $3\sigma$ sensitivity to $v_0$ for three noise levels: $c=0$ (orange, dashed), $c=2\epsilon^2\tau^2$ (green, dashed), and $c=20\epsilon^2\tau^2$ (purple, dashed). 
    (b) The right panel shows $\Mbar$ as a function of the baseline angle measured from the direction toward the Galactic center, with the baseline confined to the $(\hat{\bm X}, \hat{\bm Y})$ plane and $x=mv_0|\Delta\vec r\,|=1$ held fixed, taking $\vec{v}_{\rm obs} = \vec  v_{\rm LSR} + \vec  v_{\odot,{\rm pec}} + \vec{v}_{\oplus}({\rm March\, 20})$. We define the baseline direction by $\rhat(\theta)=\cos\theta\,\hat{\bm X}
+\sin\theta\,\hat{\bm Y}$, so that $\theta=0$ points toward the Galactic center and $\theta=\pi/2$ points along Galactic rotation. The red band shows the full annual range of $\SHMpp$ response. The right axis shows the number of measurements $N_{3\sigma}^{(v_{{\rm obs},X})}$ for $\SHMpp$ required to achieve $3\sigma$ sensitivity to $v_{{\rm obs},X}$. }
    \label{fig:M_distributions}
\end{figure*}

Before evaluating the sensitivity to different halo models, we first account for instrumental noise that can affect the measured signal. In Section~\ref{sec:observable}, we described how averaging over the dark-matter velocity distribution reduces the coherence between the two nodes. From the perspective of the two-qubit density matrix, this loss of coherence corresponds to an increase in the mixedness of the state. 
It is important, however, to distinguish this intrinsic mixedness from that induced by instrumental noise. 
The former is already encoded in the signal density matrix $\widetilde\rho$ in Eq.~\eqref{eq:density} and reflects the underlying velocity distribution we aim to probe. The latter introduces additional incoherent contributions that further suppress the measurable exchange coherence that carries directional information. To quantify this effect, we adopt the effective depolarization model of Ref.~\cite{Fukuda2025}.

We assume that the two qubits are subject to independent depolarization before postselection and nonlocal readout, with the same noise strength at both nodes. At leading order in the weak-excitation and small-noise expansion, this noise adds equal, incoherent populations to the $|10\rangle$ and $|01\rangle$ states, without contributing to the coherence between them.
Let $\rho(\tau)$ denote the ensemble-averaged two-qubit state before postselection, where the average is over field realizations, and $\rho'(\tau)$ denote the corresponding state after depolarization.
The corresponding one-excitation probabilities are 
\begin{eqnarray}
    p_1&\equiv&\Tr[\Pi_1\rho(\tau)] =P_{\rm 1exc}\simeq2\epsilon^2\tau^2,  \\
    p_1'&\equiv&\Tr[\Pi_1\rho'(\tau)].
\end{eqnarray}
The difference between these probabilities stems from the additional one-excitation events induced by the noise. 
We remark that the conditional signal state derived in Eq.~\eqref{eq:density} can also be written as
\begin{equation}
    \widetilde\rho=\frac{1}{p_1}\Pi_1\rho(\tau)\Pi_1 \, .
\end{equation}
Following the effective depolarization model in Ref.~\cite{Fukuda2025}, we introduce a small noise parameter $c\ll 1$, which defines the additional incoherent contribution to the one-excitation subspace per interrogation. 
We therefore have the corresponding block of the noisy density matrix
\begin{equation}
 \Pi_1\rho'(\tau)\Pi_1
 \simeq p_1\widetilde\rho+\frac{c}{2}\Pi_1,
 \qquad p_1'\simeq p_1+c.
 \label{eq:fukuda-noisy-state}
\end{equation}
We emphasize that this expression describes only the one-excitation subspace relevant to the postselected measurement. The populations in the remaining sectors of the full two-qubit density matrix are not shown, but must be considered to preserve the overall normalization. 

Conditioning on an accepted outcome, we find
\begin{align}
 \tilde{\rho}'
 &\equiv\frac{\Pi_1\rho'(\tau)\Pi_1}{p_1'}
 \simeq\frac{p_1\widetilde\rho+(c/2)\Pi_1}{p_1+c} \nonumber \\
 & =\Dil(\xi)\widetilde\rho+
 [1-\Dil(\xi)]\frac{\Pi_1}{2},
 \label{eq:fukuda-postselected-noise}\\
 \xi&\equiv\frac{c}{\epsilon^2\tau^2}\, ,
 \qquad
 \Dil(\xi)\equiv\frac{2}{2+\xi}.
 \label{eq:suppression}
\end{align}
Here $\xi/2\simeq c/P_{\rm 1exc}$ is the ratio of incoherent to signal-induced one-excitation probabilities, while $\mathcal{D}(\xi)$ parameterizes the fraction of accepted events attributable to the signal.  
In the ordered basis $\{|10\rangle,|01\rangle\}$,
the noisy conditional state takes the form
\begin{equation}
 \tilde{\rho}'(d,h)=\frac12
 \begin{pmatrix}
 1&\Dil(\xi)C_{\rm lab}^{*}(\vec q;d,h)\\
 \Dil(\xi)C_{\rm lab}(\vec q;d,h)&1
 \end{pmatrix}.
 \label{eq:noisy-density-matrix}
\end{equation}
The diagonal populations remain equal, while the off-diagonal coherence is reduced by $\mathcal{D}(\xi)$. 
Since $\frac{\Pi_1}{2}$ associated with the incoherent contribution has vanishing expectation value for both $M_X$ and $M_Y$, i.e., $\Tr[(\Pi_1/2)M_X]=\Tr[(\Pi_1/2)M_Y]=0$, both quadrature expectation values are reduced by $\mathcal{D}(\xi)$. 
In particular, for the $M_Y$ readout considered here, the expectation value in the noisy postselected state is 
\begin{equation}
    \mu\equiv\Tr(\tilde{\rho}' M_Y)=g\Mbar,
\end{equation} 
where $g\equiv\Dil(\xi)$ denotes the signal suppression factor.
For our numerical benchmarks, we choose $\xi=0,2,20$, which give $g=1,1/2,1/11$, corresponding to $c=0,2\epsilon^2\tau^2,20\epsilon^2\tau^2$, following Ref.~\cite{Fukuda2025}. These choices effectively model different levels of local environmental noise and state-transfer infidelity, which raise the postselection rate through incoherent events without carrying the dark-matter spatial phase. 

Again following the approach of Ref.~\cite{Fukuda2025}, we estimate the number of measurements required to determine a given velocity parameter with a specified statistical precision. For a fixed detector setup, consider $N$ independent measurements of $M_Y$ accepted after postselection, each yielding an outcome $y=\pm1$ with expectation value $\mu$.
Because $M_Y^2=1$ in the accepted one-excitation subspace, a single measurement outcome has variance $\langle M_Y^2 \rangle-\langle M_Y \rangle^2=1-\mu^2$, and the standard error of the signal expectation value is $\delta \mu=\sqrt{1-\mu^2}/\sqrt{N}$.
For a small variation $\delta a$ in a velocity parameter $a$, the corresponding change in the signal expectation value is $\delta\mu\simeq(\partial\mu/\partial a)\delta a$, with the detector configuration and all the other parameters held fixed. 
We therefore find
\begin{equation}
 \delta a=
 \frac{\sqrt{1-\mu^2}}
 {\sqrt{N}\,\lvert\partial\mu/\partial a\rvert}.
\end{equation}
We then define $N_{3\sigma}^{(a)}$ as the number of accepted measurements required to satisfy the nominal $3\sigma$ relative-precision criterion
$|a|/\delta a=3$. This yields
\begin{equation}
 N_{3\sigma}^{(a)}
 =\frac{9(1-\mu^2)}
 {\left(a\,\partial\mu/\partial a\right)^2}.
 \label{eq:n3sigma-velocity}
\end{equation}

\begin{figure*}[ht]
    \begin{center}
    \subfigure[]{\includegraphics[width=0.48\linewidth]{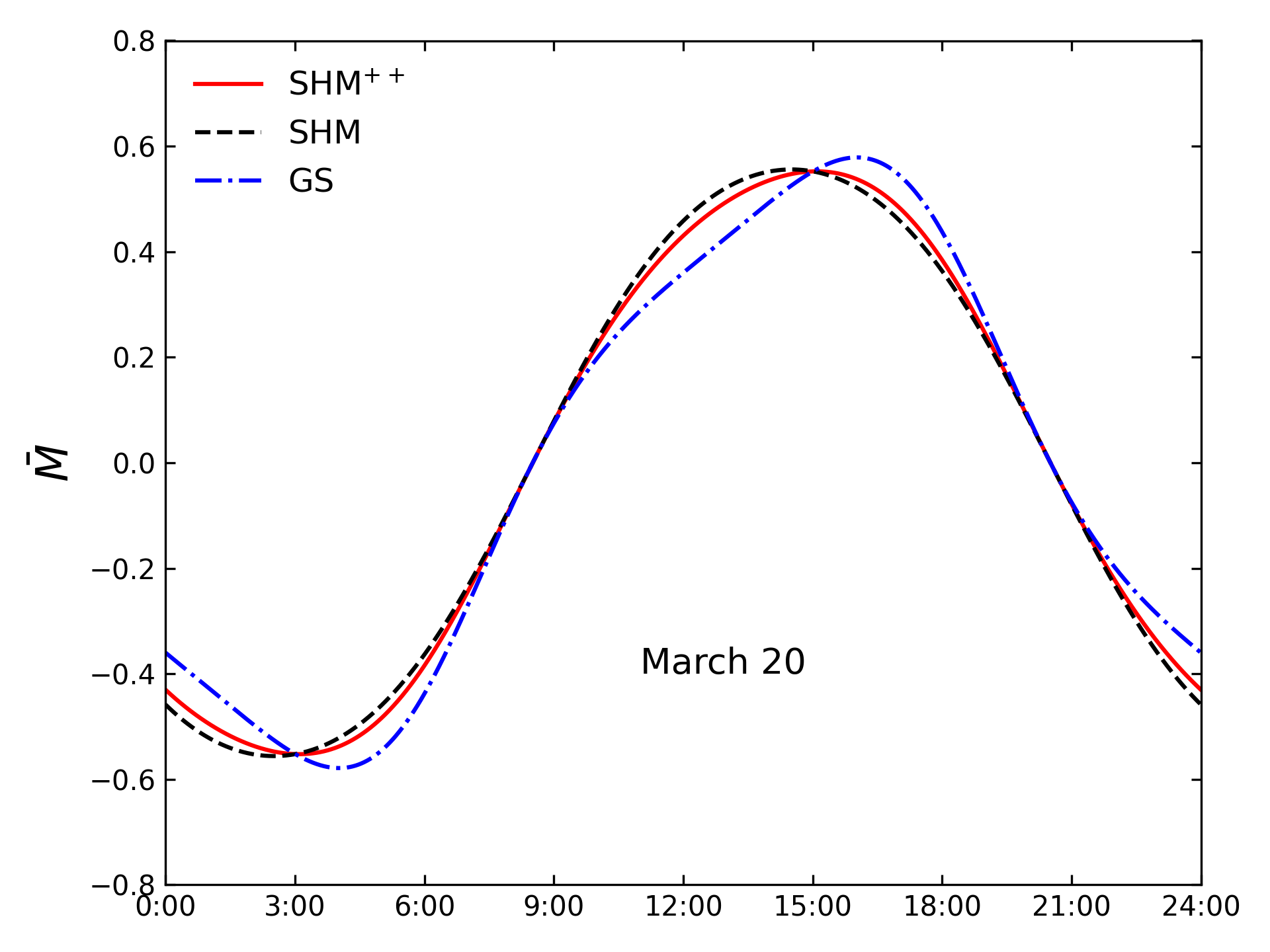}}
    \subfigure[]{\includegraphics[width=0.48\linewidth]{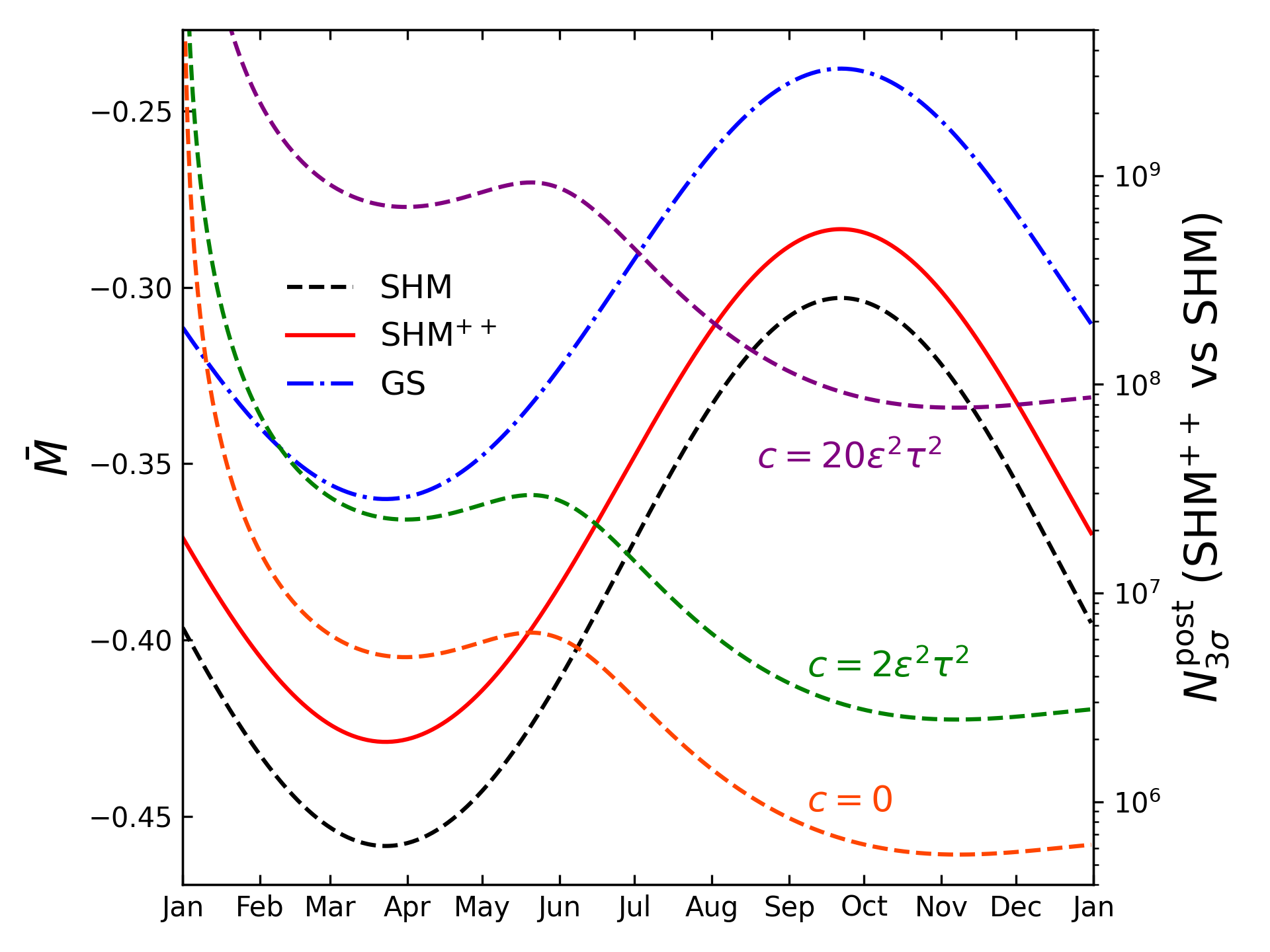}}
    \end{center}
    \vspace*{-0.6cm}
    \caption{Signal response \(\Mbar\) as a function of LST \(h\) and day of year \(d\) for an Earth-fixed east–west baseline in Lawrence, Kansas (latitude \(38.96^\circ\) N). The observer velocity \(\vec{v}_{\rm obs}\) includes the LSR, solar peculiar, orbital, and terrestrial rotational contributions. (a) Sidereal modulation of \(\Mbar\) on March 20. The solid red, dash-dotted blue, and dashed black curves correspond to \(\SHMpp\), Gaia Sausage, and SHM, respectively. (b) Annual modulation of \(\Mbar\) at fixed LST \(h=0\) (left \(y\)-axis), together with the total number of postselected measurements \(N_{3\sigma}^{\rm post}\) required to reject the SHM at \(3\sigma\), assuming \(\SHMpp\) as the true velocity distribution (right \(y\)-axis). The dashed orange, green, and purple curves correspond to depolarization levels \(c=0\), \(c=2\epsilon^2\tau^2\), and \(c=20\epsilon^2\tau^2\), respectively. The $N_{3\sigma}^{\rm post}$ curves assume uniform accepted exposure at $h=0$ from January~1 through each plotted date. For each interval, one common offset $b$ is profiled and $g=\Dil(\xi)$ is fixed to its calibrated value, using the likelihood analysis in Section~\ref{sec:statistics}.   }
    \label{fig:Mpp_n3}
\end{figure*}

We first present two idealized design scans in Fig.~\ref{fig:M_distributions}  that explore how the directional response and the statistical precision of individual velocity parameters depend on the baseline length and its orientation in the Galactic frame, varying one while holding the other fixed.
We compare the $\Mbar$ responses of the SHM, Gaia Sausage, and $\SHMpp$ on the left $y$-axes. The right $y$-axes show the required measurement counts for $\SHMpp$ by applying the local parameter-precision criterion in Eq.~\eqref{eq:n3sigma-velocity}. For illustration, we use ($\eta, \beta$) = $(0.3, 0.9)$ as mentioned earlier. Here the baseline detector is assumed to remain fixed in the Galactic frame, without accounting for Earth's rotation. For the angular scan, we parameterize the baseline direction as $\rhat(\theta)=\cos\theta\,\hat{\bm X}+\sin\theta\,\hat{\bm Y}$.  
Fig.~\ref{fig:M_distributions}(a) shows $\Mbar$ as a
function of $x=mv_0|\Delta\vec r|$, varying the baseline length while holding the mass and fiducial velocity $v_0=238\ \mathrm{km\,s^{-1}}$ fixed. The baseline points along Galactic $+\hat{\bm Y}$, and we adopt the constant observer velocity $\vec v_{\rm obs}=v_0\hat{\bm Y}$ for this idealized scan.
For the right-axis sensitivity, varying $v_0$ changes
both the halo width and the assumed observer speed
through $\vec v_{\rm obs}=v_0\hat{\bm Y}$, while the
physical baseline and other independent parameters
are held fixed.
Fig.~\ref{fig:M_distributions}(b) shows instead $\Mbar$ as a function of the baseline direction angle $\theta$ on the $({\hat{\bm X}, \hat{\bm Y}})$ plane while maintaining $x = 1$ and $\vobs = \vec  v_{\rm LSR} + \vec  v_{\odot,{\rm pec}} + \vec{v}_{\oplus}({\rm March\, 20})$. The parameter chosen for the relative-precision is $v_0$ for Fig.~\ref{fig:M_distributions}(a) and $\vobs \cdot \hat{{\bm X}}$ for Fig.~\ref{fig:M_distributions}(b). All fixed-date curves in panel (b) except the two dotted curves are calculated using untruncated velocity distributions, with measurements assumed to take place on March~20. The black and red dotted curves show the corresponding SHM and $\SHMpp$ responses with an escape-speed cutoff of $v_{\rm esc}=544\ \mathrm{km\,s^{-1}}$. The cutoff produces only small changes in $\Mbar$ for these benchmarks, motivating the use of untruncated distributions in the subsequent analysis.

We next show the sidereal and annual modulation of the signal response $\overline M$ for an Earth-fixed apparatus in Fig.~\ref{fig:Mpp_n3}, with an east-west baseline located at Lawrence, Kansas (latitude $38.96^{\circ}$). 
The response is shown as a function of LST $h$ and day of year $d$.
In addition, the right $y$-axis of panel (b) shows the total number of accepted measurements required to reject the SHM at the $3\sigma$ level, assuming the benchmark $\SHMpp$ model is the true velocity distribution. The required exposure is calculated using Eq.~\eqref{eq:n3sigma}, with the statistical procedure detailed in Section~\ref{sec:statistics}.

\begin{figure*}[t]
    \begin{center}
    \subfigure[]{\includegraphics[width=0.48\linewidth]{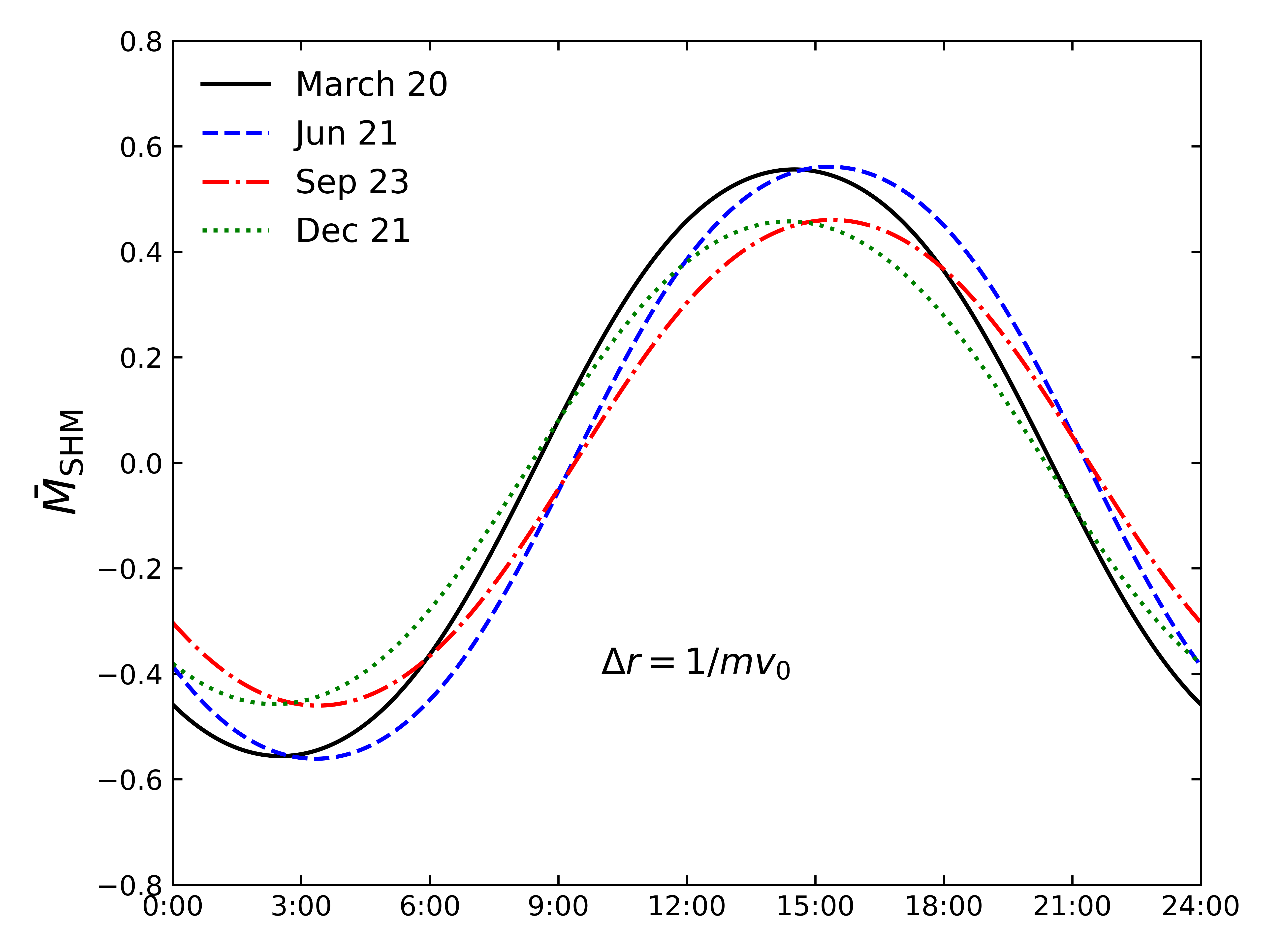}}
    \subfigure[]{\includegraphics[width=0.48\linewidth]{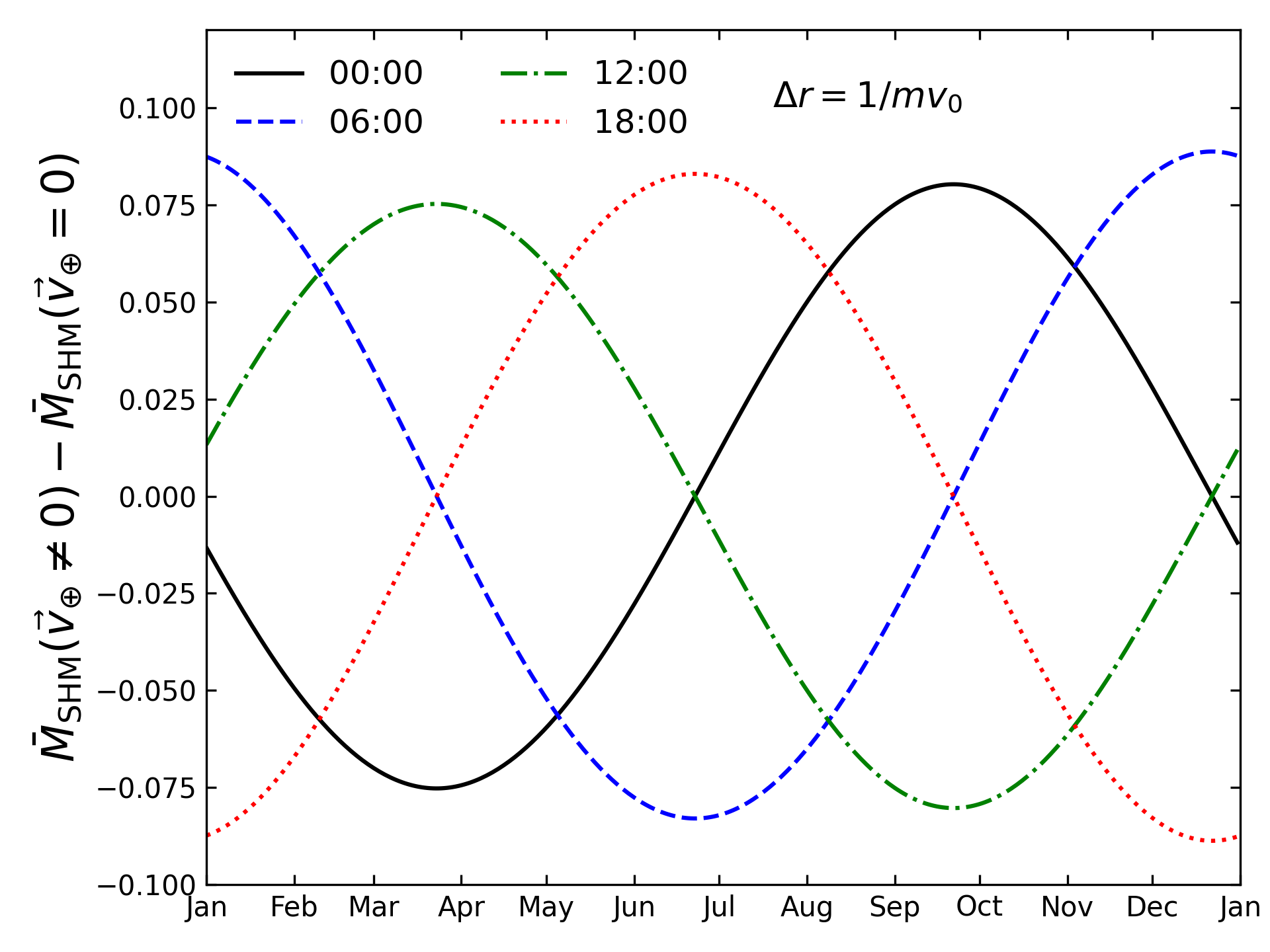}}
    \end{center}
    \vspace*{-0.6cm}
    \caption{Modulation of the noise-free, untruncated SHM response for an Earth-fixed two-node baseline at $x=mv_0 |\Delta\vec r\,|=1$. (a) Sidereal response versus LST on four dates. The baseline remains fixed locally while its direction rotates in Galactic coordinates. (b) Exact annual residual at four fixed sidereal phases, obtained by subtracting a modeled no-orbit response while retaining the same baseline orientation, LSR and solar peculiar velocities, and laboratory rotational velocity, i.e., the difference between $\Mbar$ with and without the annual modulation. The baseline is horizontal and directed eastward in Lawrence, Kansas, at latitude $38.96^\circ$ N.}
    \label{fig:earth-fixed}
\end{figure*}

Fig.~\ref{fig:earth-fixed} examines the Earth-fixed geometry used in Fig.~\ref{fig:Mpp_n3} in more detail. The two nodes and their directed separation remain fixed. As the apparatus (i.e., the two-node system) rotates with Earth, $|\Delta\vec r\,|$ and its local orientation remain constant, while its direction $\rhat(h)$ changes in equatorial and Galactic coordinates. Fig.~\ref{fig:earth-fixed}(a) evaluates the SHM response over one sidereal day on four representative dates, using the full observer velocity $\vobs(d,h)$ in Eq.~\eqref{eq:vobs}, which includes $\vec{ v}_{\rm LSR}$, the Sun's peculiar velocity, Earth's orbital velocity, and the laboratory velocity due to Earth's rotation.
The last contribution $\vrot(h)$ has a magnitude of approximately $0.36\ \mathrm{km\,s^{-1}}$ at this latitude. For the eastward baseline, $\vrot(h)$ is parallel to $\rhat(h)$, so its projection onto the baseline remains constant throughout a sidereal day. The daily variation instead comes from the changing projection of the other observer-velocity terms onto $\rhat(h)$. This complements Fig.~\ref{fig:Mpp_n3}(a), which compares three velocity models on a single date.
By contrast, Fig.~\ref{fig:earth-fixed}(b) shows the exact annual residual at four fixed LST values $h=0,6,12,$ and $18$ hours. At each fixed $h$, the baseline direction and $\vrot(h)$ remain unchanged throughout the year, while Earth's orbital velocity $\vorb(d)$ introduces the annual time dependence. The residual is defined as the difference between the full response and a reference prediction obtained by setting \(\vorb=0\), with all other contributions held fixed. 
Although the full response can be measured at the same LST on different days using an Earth-fixed apparatus, the reference prediction without orbital motion is counterfactual and cannot be measured independently. Fig.~\ref{fig:earth-fixed}(b) thus complements the full annual response curves in Fig.~\ref{fig:Mpp_n3}(b).

At fixed LST $h$, the broadband response in Eq.~\eqref{eq:response-x} factorizes into a common day-dependent sine factor and a model-dependent coherence amplitude $A_{\mathcal H}(x,\rhat(h))$, which remains constant throughout the year. 
The centered, even halo models considered here therefore share the same annual waveform, differing only in their overall amplitudes, as illustrated in
Fig.~\ref{fig:Mpp_n3}(b). For two halo hypotheses $a$ and $b$ with $A_b\ne0$, we have
\begin{equation}
 \Mbar_a(d;h)
 =\frac{A_a(x,\rhat(h))}{A_b(x,\rhat(h))}
  \Mbar_b(d;h).
 \label{eq:fixed-lst-ratio}
\end{equation}
The same proportionality holds for their annual residuals. Consequently, if the effective signal amplitude is fitted freely and the required rescaling lies within its allowed range, the SHM and centered $\SHMpp$ predictions become 
indistinguishable at a single LST, even in the presence of a common constant readout offset.
Although Earth's orbital motion changes their shared phase waveform, it cannot break this amplitude degeneracy. The fixed-LST discrimination result in Fig.~\ref{fig:Mpp_n3}(b) therefore 
assumes a calibrated effective signal amplitude. 

On the other hand, joint annual--sidereal observations can break this degeneracy by probing different baseline directions as Earth rotates. While the SHM coherence amplitude depends only on $x = m v_0 |\Delta\vec r\,|$, the anisotropic $\SHMpp$ amplitude also depends on $\rhat(h)$. 
For the east--west baseline considered here, their amplitude ratio therefore varies with LST. 
A single global rescaling cannot reproduce this variation, allowing the models to be distinguished when the signal scale is common to all observing times and the readout offset is stable or independently constrained. 
However, the degeneracy persists if the sampled baseline directions yield the same amplitude ratio.

\begin{figure*}[t!]
    \begin{center}
    \subfigure[]{\includegraphics[width=0.48\linewidth]{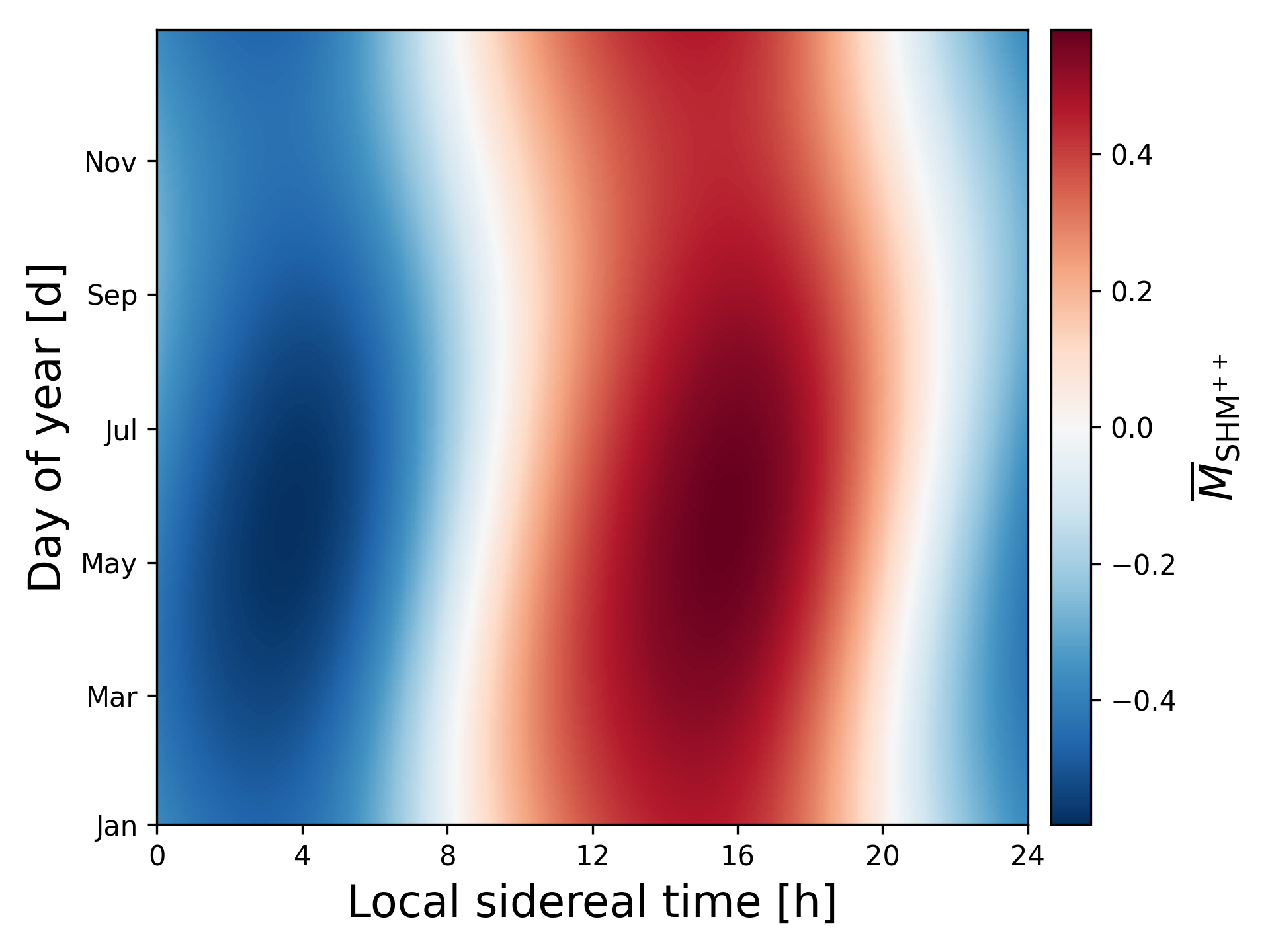}}
    \subfigure[]{\includegraphics[width=0.48\linewidth]{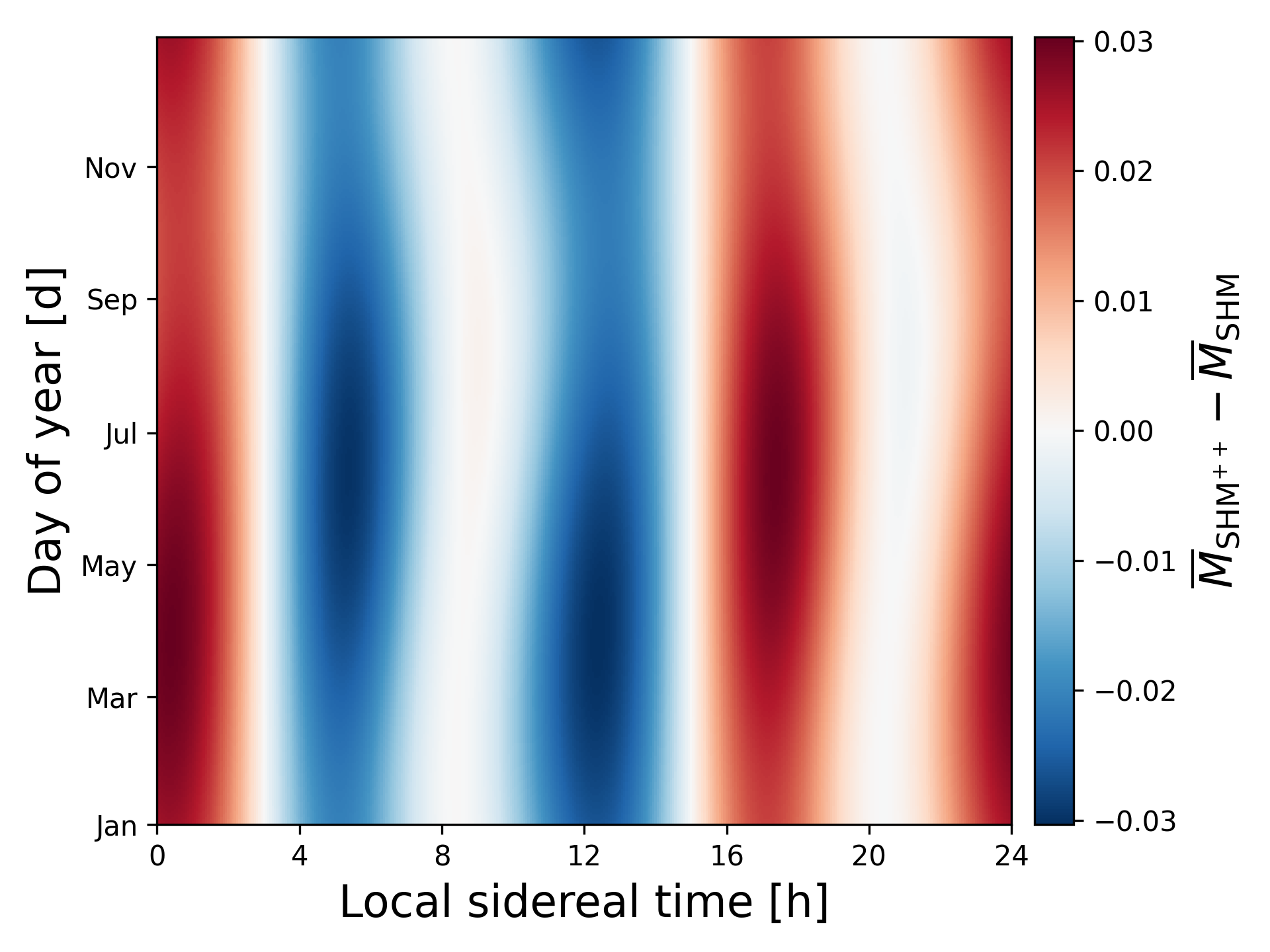}}
    \end{center}
    \vspace*{-0.6cm}
    \caption{Joint annual--sidereal response for $c = 0$: $\Mbar_{\SHMpp}$ in the left (a) and the difference between $\Mbar_{\SHMpp}$ and $\Mbar_{\rm SHM}$ in the right (b). The $y$ axis is the day of the year and the $x$ axis is the local sidereal time.}
    \label{fig:M_color}
\end{figure*}

Fig.~\ref{fig:M_color} further illustrates the joint annual--sidereal structure over the $(d,h)$ plane in the absence of depolarization $c=0$. The left panel shows $\Mbar_{\SHMpp}$, while the right panel shows the raw difference $\Mbar_{\SHMpp}-\Mbar_{\rm SHM}$ before fitting an offset or rescaling the signal amplitude. To quantify the statistical significance of these differences, we next perform a likelihood-based analysis that accounts for the unknown signal amplitude and readout offset. 

\section{Discrimination of SHM and $\SHMpp$}
\label{sec:statistics}

We now quantify the statistical sensitivity to distinguishing different halo models using the modulation patterns discussed in the previous section. To account for possible systematic effects, we model the measured expectation value of the postselected binary outcomes in each observation bin $k$ as 
\begin{equation}
 \mu_k=b+g\Mbar_{\mathcal H}(d_k,h_k),
 \qquad |\mu_k|<1,
 \label{eq:mean-model}
\end{equation}
where $b$ denotes a stable additive readout asymmetry; it is a readout offset, neither the incoherent-event probability $c$ nor a background event rate. The factor $g$ is an effective response amplitude shared across the bins; in the calibrated analysis $g=\Dil(\xi)$, whereas in a profiled analysis, $g$ is treated as a fitting parameter that may also absorb an overall 
readout gain uncertainty. 
Throughout this analysis, the postselected exposure $N_{\rm post}$ includes all accepted coherent and incoherent events.

Since $M_Y$ has eigenvalues $\pm1$ in the accepted subspace, its predicted mean $\mu_k$ determines the probabilities of both outcomes in each observation bin $k$:
\begin{equation}
 p_{k,+}=\frac{1+\mu_k}{2},
 \qquad
 p_{k,-}=\frac{1-\mu_k}{2}.
 \label{eq:binary-probabilities}
\end{equation}
Denoting the total number of measurements accepted after postselection for a given bin $k$ by $N_k$ and the numbers of measurements yielding $M_Y=\pm 1$ by $n_{k,\pm}$, we have  $N_k=n_{k,+}+n_{k,-}$.
Assuming independent measurement outcomes, the log-likelihood, up to binomial terms independent of $(\mathcal H,b,g)$, is given by
\begin{equation}
 \ln\mathcal L(\mathcal H,b,g)=\sum_k
 \left[n_{k,+}\ln p_{k,+}+n_{k,-}\ln p_{k,-}\right].
 \label{eq:likelihood}
\end{equation}
We condition on the number of accepted events in each bin and use their $M_Y=\pm1$ outcomes. The likelihood therefore excludes the number of interrogations that fail postselection. 
This treatment is justified at leading order in the matched, broadband-response limit, where the signal-induced acceptance  probability $2\epsilon^2\tau^2$ is independent of the halo shape. A velocity-dependent response can also make the acceptance rate sensitive to the halo model. This information can be included through an additional likelihood for accepted and rejected event counts.

We now estimate the number of accepted measurements required to reject the SHM, assuming that the benchmark $\SHMpp$ distribution is the true model.  
In the following, $T$ labels the assumed true model, while $H$ labels the tested SHM hypothesis. 
To quantify the statistical difference between their predicted binary outcome distributions with means $\mu_T$ and $\mu_H$, we use the Kullback–Leibler (KL) divergence,
\begin{equation}
 \DKL(\mu_T\Vert\mu_H)=
 \frac{1+\mu_T}{2}\ln\frac{1+\mu_T}{1+\mu_H}
 +\frac{1-\mu_T}{2}\ln\frac{1-\mu_T}{1-\mu_H}.
 \label{eq:kl}
\end{equation}
We define $w_k=N_k/N_{\rm post}$ as the fraction of accepted measurements assigned to bin $k$, with $N_{\rm post}=\sum_kN_k$ being the total number of accepted measurements.
For a representative data set, we set each outcome count equal to its expectation value under the assumed true model:
\begin{equation}
 \bar n_{k,\pm}
 \equiv\mathbb E_T[n_{k,\pm}\mid N_k]
 =N_k\frac{1\pm\mu_k^{(T)}}{2}.
\end{equation}
These expected counts need not be integers.

\begin{table*}[t]
\centering
\caption{Benchmark inputs for the likelihood analysis.}
\label{tab:benchmark}
\begin{tabular}{@{}lc|cl@{}}
\toprule
Quantity &&& Value \\
\hline
\midrule
Circular speed, halo width &&& $v_c=v_0=238\ \mathrm{km\,s^{-1}}$ \\
Site &&& Lawrence, Kansas; latitude $38.96^\circ$ N \\
Directed baseline &&& horizontal, directed eastward \\ 
Dimensionless separation &&& $x=m v_0 |\Delta\vec r\,| =1$ \\
$\SHMpp$ truth &&& $\eta=0.3$, $\beta=0.9$ \\
Joint annual--sidereal schedule &&& 24 annual phases $\times$ 24 LST phases \\
Exposure &&& equal accepted-event exposure per bin \\
Nuisance parameters &&& constant offset $b$; effective amplitude $g$ where indicated \\
\bottomrule
\end{tabular}
\end{table*}

The one-sided profile-likelihood statistic compares the SHM fit
with the best fit in the mixture family:
\begin{equation}
 q_0=-2\ln
 \frac{\displaystyle\max_{\bm\nu}
       \mathcal L(\eta=0,\bm\nu)}
      {\displaystyle\max_{0\leq\eta\leq1,\,\bm\nu}
       \mathcal L(\eta,\bm\nu)}.
\end{equation}
Here $\mathcal L(\eta,\bm\nu)$ is the likelihood in Eq.~\eqref{eq:likelihood} evaluated for the mixture model with Sausage fraction $\eta$. The nuisance set is $\bm\nu=\{b\}$ when the effective amplitude is calibrated and $\bm\nu=\{b,g\}$ when it is profiled. The nuisance parameters are common to all bins. For all estimates of $N_{3\sigma}^{\rm post}$, we take $b_T=0$ and $g_T=\Dil(\xi)$ for the assumed true model.
The anisotropy parameter $\beta$ is held fixed in each likelihood-ratio test.

\begin{figure}[t]
\centering
    \includegraphics[width=0.48\textwidth]{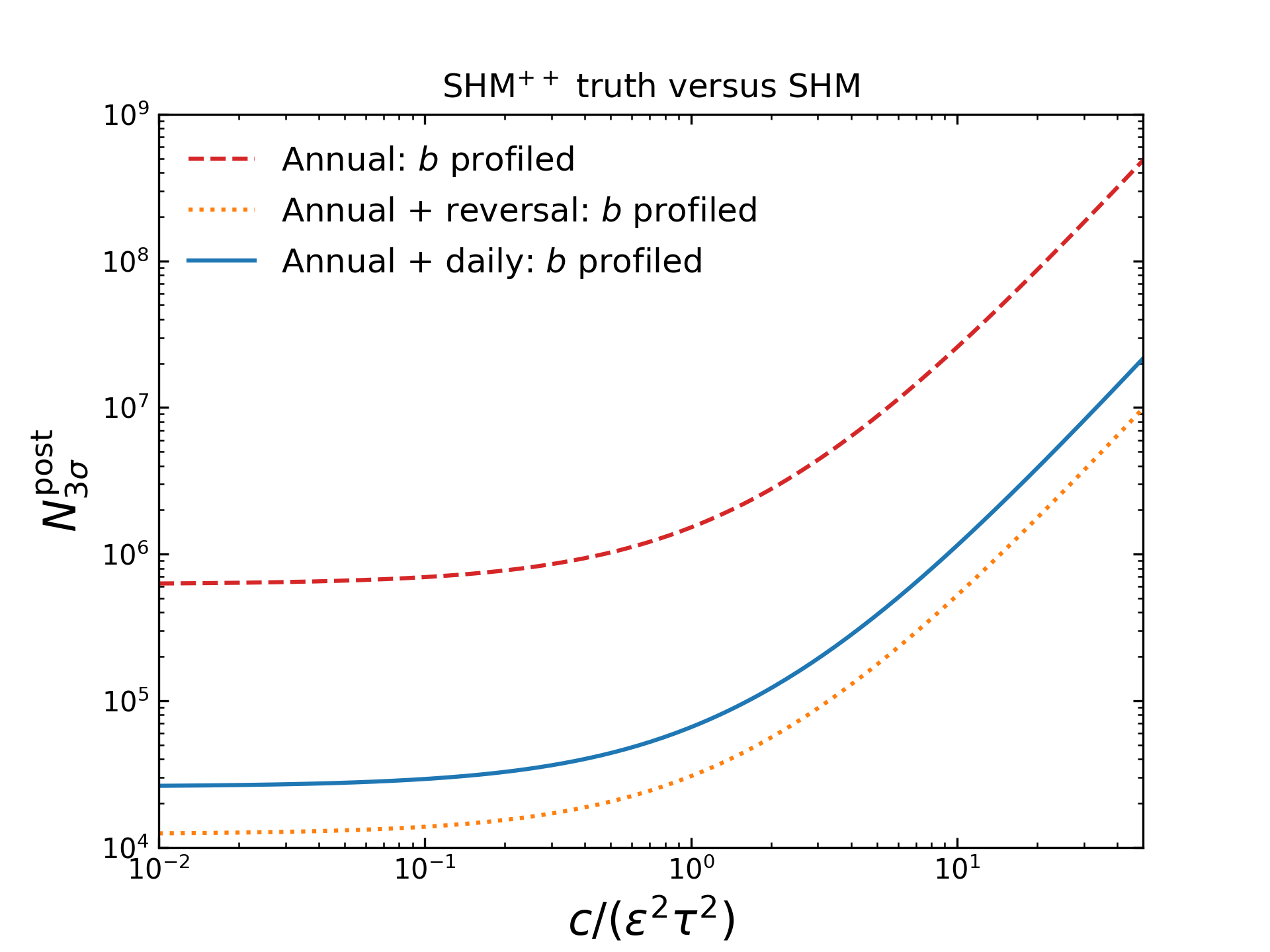}
    \caption{Number of measurements $N_{3\sigma}^{\rm post}$ required for $3\sigma$ rejection of the SHM, assuming the $\SHMpp$ to be the true model, as a function of the depolarization ratio $c/(\epsilon^2\tau^2)$. The effective signal amplitude $g$ is fixed to its calibrated value.}
    \label{fig:N_noise}
\end{figure}

\begin{figure*}[t]
 \centering
    \subfigure[]{\includegraphics[width=0.48\linewidth]{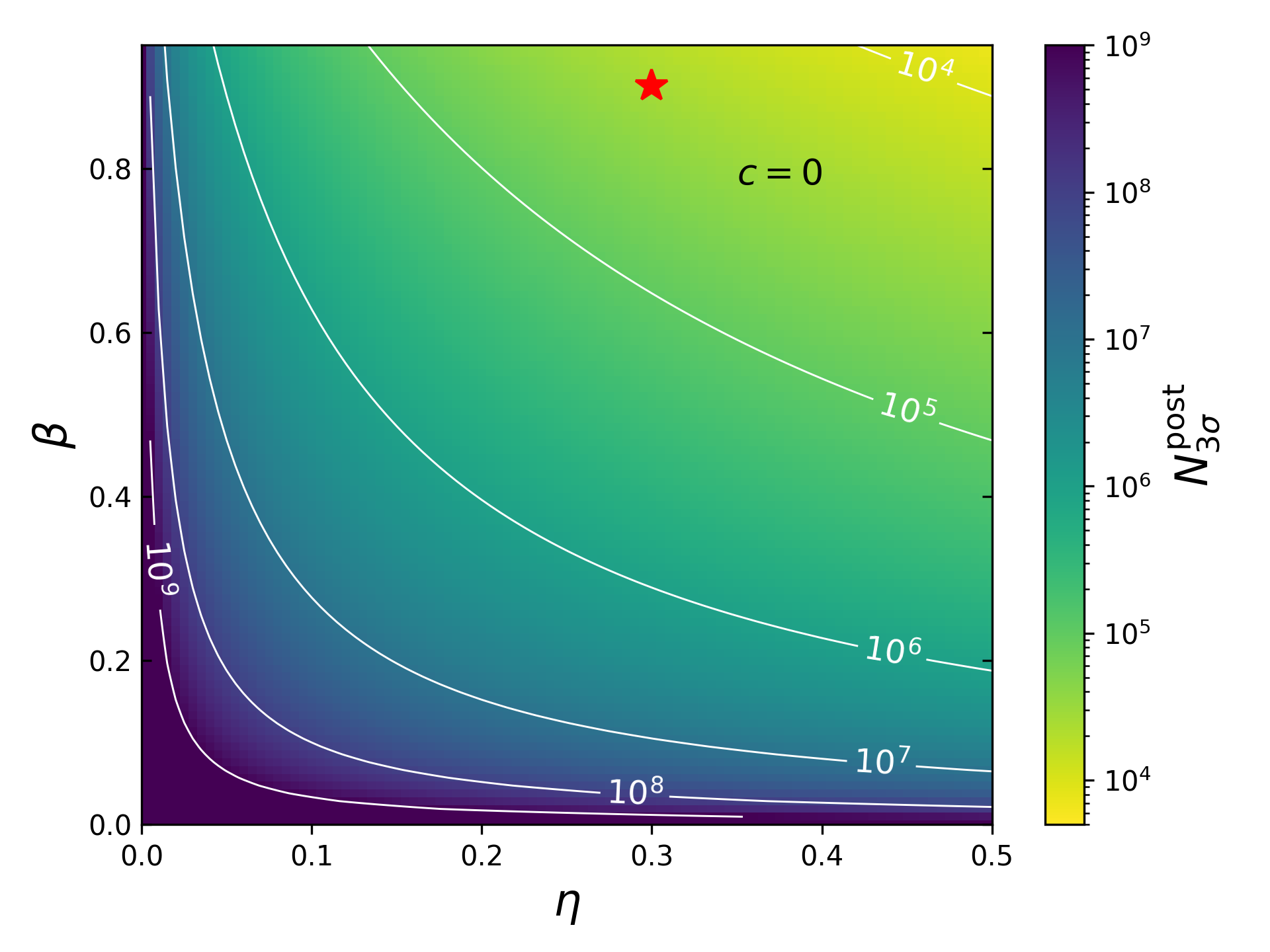}}
    \subfigure[]{\includegraphics[width=0.48\linewidth]{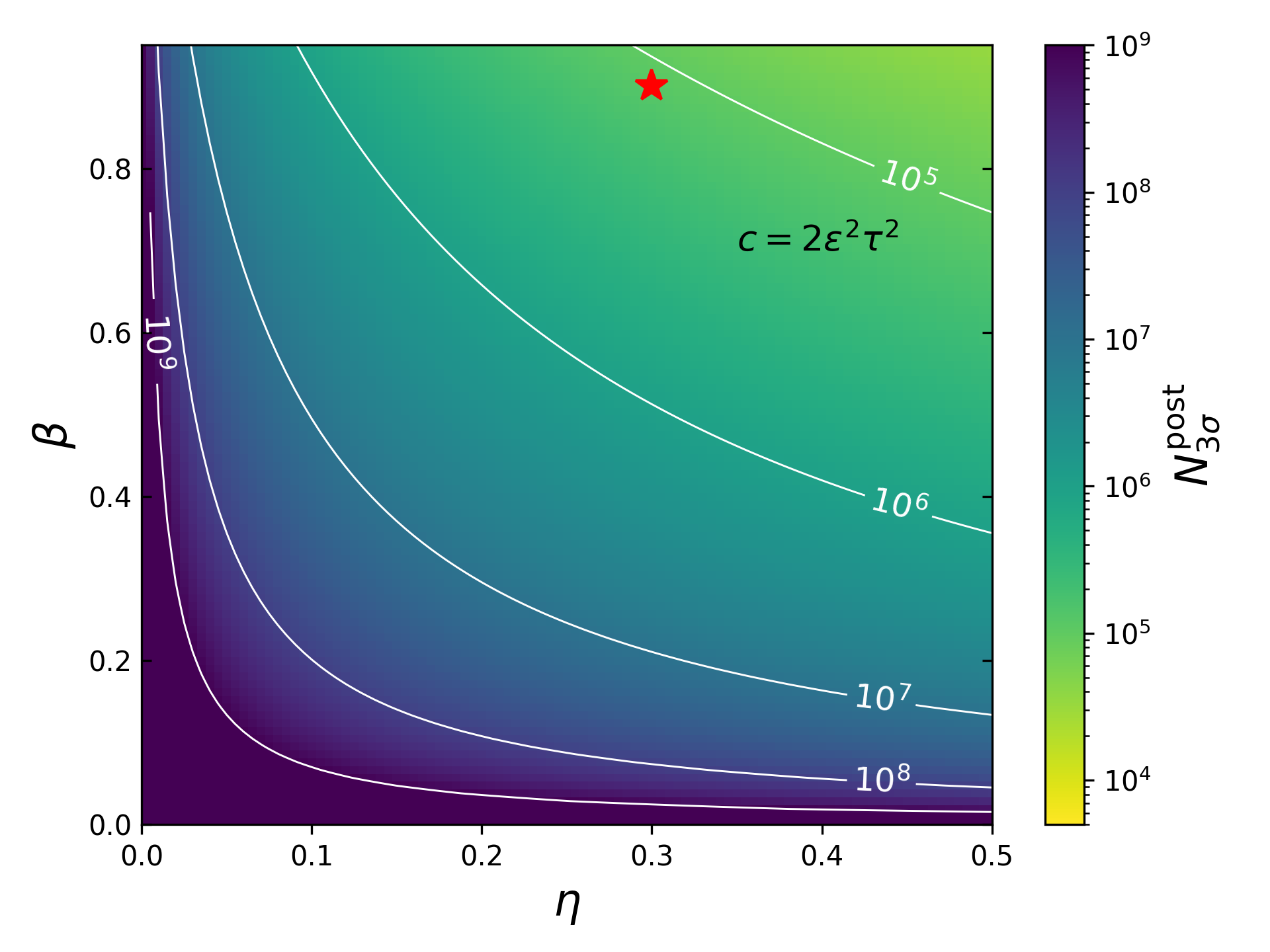}}
    \vspace*{-0.3cm}
    \caption{Required accepted exposure in the $(\eta,\beta)$ plane for an $\SHMpp$ truth against the SHM.  The constant offset and effective global amplitude are profiled in every pixel.  Panels (a) and (b) show $\xi=0$ and $\xi=2$. White contours give constant $N_{3\sigma}^{\rm post}$. The star marks the benchmark point $(\eta,\beta)=(0.3,0.9)$.}
 \label{fig:eta-beta}
\end{figure*}

For the representative data, the denominator is maximized at the assumed true model $T$, which was used to generate the expected outcome counts. 
Substituting the expected counts $\bar n_{k,\pm}$ into Eq.~\eqref{eq:likelihood} therefore gives
\begin{align}
 q_{0,\mathrm{rep}}
 &=2\left[
 \max_{0\leq\eta\leq1,\,\bm\nu}
     \ln\mathcal L(\eta,\bm\nu)
 -\max_{\bm\nu}
     \ln\mathcal L(\eta=0,\bm\nu)
 \right] , \nonumber \\ 
 &=2\min_{\bm\nu}\sum_{k=1}^{K}N_k \DKL\!\left[      \mu_k^{(T)}\Vert\mu_k^{(H)}(\bm\nu) \right], \\
   &=N_{\rm post}\, \mathcal I_{T:H}, \nonumber  \\
 \mathcal I_{T:H}
 &\equiv2\min_{\bm\nu}\sum_{k=1}^{K}w_k
   \DKL\!\left[
     \mu_k^{(T)}\Vert\mu_k^{(H)}(\bm\nu)
   \right],
 \label{eq:asimov-info}
\end{align}
where $\mathcal I_{T:H}$ is the profiled likelihood-ratio information per accepted measurement for the chosen observing schedule. 
Equal accepted exposure in every bin gives $w_k=1/K$.
For the calculations in Figs.~\ref{fig:N_noise} and~\ref{fig:eta-beta}, the annual phase samples are $d_j=1+365j/24$, with $j=0,\ldots,23$. The fixed-LST analysis uses $h=0$, while the joint analysis additionally samples $h_\ell=\ell\,\mathrm{h}$, with $\ell=0,\ldots,23$. Responses are evaluated at these phase points without averaging over finite time bins. The nuisance fit allows $b\in\mathbb R$ and, when profiled, $g\geq0$, subject to $|\mu_k|<1$ in every bin; no additional bounds are imposed.

When $\eta$ remains identifiable after nuisance profiling, the one-sided asymptotic approximation gives a median significance $Z_{\rm med}\simeq\sqrt{q_{0,\mathrm{rep}}}$~\cite{Cowan2011}. We define the corresponding asymptotic median $3\sigma$ accepted-event exposure by
\begin{equation}
 N_{3\sigma}^{\rm post}
 =\frac{9}{\mathcal I_{T:H}}.
 \label{eq:n3sigma}
\end{equation}
$N_{3\sigma}^{\rm post}$ should not be confused with $N_{3\sigma}^{(a)}$; $N_{3\sigma}^{\rm post}$ quantifies the accepted exposure required for halo-model rejection, whereas $N_{3\sigma}^{(a)}$ quantifies the accepted exposure required to achieve the specified local precision on a parameter.

We first show the number of postselected measurements $N_{3\sigma}^{\rm post}$ required for a $3\sigma$ rejection of the SHM in Fig.~\ref{fig:N_noise}, assuming the benchmark $\SHMpp$ model is true, as a function of the depolarization ratio $c/(\epsilon^2\tau^2)$. All three analyses use the same physical two-node baseline, with benchmark inputs summarized in Table~\ref{tab:benchmark}. The red dashed curve uses 24 annual-phase bins at fixed
LST $h=0\,\mathrm{h}$. The blue solid curve combines annual and sidereal modulation, using 24 bins for each phase and thus $24\times24=576$ observing bins. The orange dotted curve, labeled ``reversal,'' supplements each fixed-LST annual bin with a separately acquired signal-odd setting. A physical reversal of the directed baseline, for example, changes the response according to $\Mbar(-m  \Delta\vec r)=-\Mbar(m \Delta\vec r)$. Measuring both settings at each of the 24 annual phases gives 48 phase--setting bins. Provided the readout offset is unchanged between settings, their half-difference isolates the signal while their half-sum constrains the offset. This additional information substantially reduces the required measurement counts for this benchmark. Exposure is distributed equally among the sampled bins and settings, and $N_{3\sigma}^{\rm post}$ includes accepted measurements from both reversal settings.

We next treat $(\eta,\beta)$ as the true $\SHMpp$ parameters and test the SHM at every grid point.  For each point, the annual--sidereal response is evaluated in the same $24\times24$ schedule, and the SHM likelihood is maximized over a constant $b$ and global $g$.  The resulting $N_{3\sigma}^{\rm post}$ is shown in Fig.~\ref{fig:eta-beta}.  
The scan uses a uniform $101\times101$ grid, $\eta_i=0.005i$ and $\beta_j=0.0095j$, with $i,j=0,\ldots,100$. On either exact degeneracy axis, $\eta=0$ or $\beta=0$, the required exposure is $N_{3\sigma}^{\rm post}=\infty$. For display only, values above $10^9$, including infinities, are assigned the upper-limit color.
This is a model-rejection map, not a two-parameter confidence region around an estimator.

The two exact degeneracy axes have a simple interpretation.  If $\eta=0$, the mixture contains only the SHM.  If $\beta=0$, Eq.~\eqref{eq:variances} gives $a_r=a_t=1/2$, so the nominal Sausage component is itself identical to the SHM.  No amount of exposure can distinguish identical distributions. Expanding Eq.~\eqref{eq:variances} near the SHM point $(\eta,\beta)=(0,0)$, we obtain 
\begin{equation}
 a_r=\frac12+\frac{\beta}{3}+\mathcal O(\beta^2),
 \qquad
 a_t=\frac12-\frac{\beta}{6}+\mathcal O(\beta^2).
 \label{eq:small-beta-variances}
\end{equation}
For a unit direction $\rhat$,
\begin{equation}
 a_r\hat r_X^2+a_t(\hat r_Y^2+\hat r_Z^2)
 =\frac12+\frac{\beta}{6}(3\hat r_X^2-1)+\mathcal O(\beta^2).
 \label{eq:small-beta-projection}
\end{equation}
Consequently, the template departure satisfies
$\Mbar_{++}-\Mbar_{\rm SHM}=\mathcal O(\eta\beta)$.  Since the KL divergence is quadratic in a small template difference after nuisance projection,
\begin{equation}
 \mathcal I_{++:{\rm SHM}}\propto(\eta\beta)^2,
 \qquad
 N_{3\sigma}^{\rm post}\propto(\eta\beta)^{-2},
 \label{eq:small-departure-scaling}
\end{equation}
up to geometry- and likelihood-dependent corrections (see Appendix \ref{sec:degeneracy} for more details.).  The bending of the numerical contours away from the axes reflects higher orders and the finite range of $|\mu|$.

These results show that joint annual--sidereal measurements can distinguish an anisotropic $\SHMpp$ distribution from the SHM through response differences that survive profiling a constant offset and a global amplitude. Depolarization increases the required accepted exposure, while approaching
either SHM limit progressively removes the distinguishing
signal. The analysis therefore quantifies the discrimination
sensitivity of the joint annual--sidereal measurements
and identifies the exact degeneracies that additional
exposure cannot resolve.

Moving from model discrimination to a measurement of the velocity distribution requires broader Fourier-space coverage. With calibrated readout offsets and response amplitudes, measurements of both exchange quadratures give
$ C_{\rm lab}(\vec q;d,h)=\langle M_X\rangle-\ii\langle M_Y\rangle$. 
Removing the known laboratory phase in Eq.~\eqref{eq:shift} then determines $\Phi(\vec q)$. Varying $|\Delta\vec r|$ samples different Fourier radii, while additional baseline orientations, including directions outside the plane swept out by the east--west baseline, provide the angular coverage needed for a general three-dimensional reconstruction. 
With complete, noiseless Fourier-space coverage, the Galactic
velocity distribution is formally recovered through
\begin{equation}
 f_{\rm gal}(\vec u)=\frac{1}{(2\pi)^3}
 \int\dd^3q\,\Phi(\vec q)e^{-\ii\vec q\cdot\vec u}.
 \label{eq:fourier-inversion}
\end{equation}
In practice, measurements cover only a finite, noisy region of Fourier space, so reconstruction requires regularization and treatment of the detector response. The resulting reconstruction has finite resolution; one fixed rotating baseline does not provide full three-dimensional tomography.

We used Lawrence, KS as a benchmark site. Quantitative sensitivities may vary with the detector location and baseline geometry. The fixed-LST amplitude degeneracy persists, while joint annual--sidereal measurements retain their qualitative advantage whenever the sampled baseline directions yield different ratios of the SHM and $\SHMpp$ coherence amplitudes.

\section{Conclusions}
\label{sec:conclusions}

We have developed a framework for probing the local dark-matter velocity distribution with a two-node quantum sensor network. For two sensors with identical calibrated responses that are approximately constant across the dark-matter signal bandwidth, nonlocal exchange measurements conditioned on the one-excitation subspace determine the laboratory-frame velocity characteristic function at $\vec q=m\Delta\vec r$. We applied this framework to the SHM, a radially anisotropic Gaia-Sausage component, and their $\SHMpp$ mixture, incorporating the annual and sidereal motion of an Earth-fixed baseline, depolarization-induced dilution, and an additive readout offset. 

A central result is the amplitude degeneracy of annual measurements at one fixed LST $h$. For the stationary, centered, even halo distributions considered here, the pure boost to $\vec{v}_{\rm obs}$ produces a common annual phase dependence, while the Galactic coherence amplitudes remain constant at fixed $h$. The two model responses are therefore proportional. If the effective response amplitude is freely profiled and the required rescaling lies within its allowed range, an annual trace alone cannot distinguish the SHM from the $\SHMpp$ benchmark. For the Earth-fixed east--west baseline studied here, however, the ratio of the two coherence amplitudes varies with $h$. A joint annual--sidereal analysis consequently retains directional information that cannot be absorbed by one common offset and one global response amplitude.

For the untruncated benchmark $x=mv_0 |\Delta\vec r\,|=1$ and
$(\eta,\beta)=(0.3,0.9)$, we evaluated the full Bernoulli likelihood on a $24\times24$ annual--sidereal schedule with equal accepted exposure per bin. 
Profiling the constant offset $b$ while fixing $g=\Dil(\xi)$, the fixed-$\beta$ test in Fig. \ref{fig:N_noise} gives asymptotic median $3\sigma$ accepted-event requirements of $N_{3\sigma}^{\rm post}=2.6\times10^4$ for $c=0$ and $1.2\times10^5$ for $c=2\epsilon^2\tau^2$. These counts include both signal and incoherent accepted events. The $(\eta,\beta)$ scan in Fig.~\ref{fig:eta-beta} is obtained by profiling both the constant offset $b$ and the
global effective amplitude $g$. It exhibits exact degeneracy on either SHM axis, $\eta=0$ or $\beta=0$. The statistical model distinguishes two effects of stationary instrumental backgrounds. A stable additive readout offset contributes to the constant mean, whereas accepted incoherent events suppress the modulation through the dilution factor. The offset can be profiled across the observing schedule or constrained by independently implemented signal-odd settings, provided calibration verifies that the offset is the same in both settings. These results motivate observations over multiple LSTs and annual phases, interleaved phase cycling, and a joint likelihood for the signal, calibration, and control data. Time-dependent backgrounds require additional constrained templates, with their overlap with the predicted signal propagated into the inferred sensitivity.

The astrophysical forecasts remain conditional on the adopted halo models. The $\SHMpp$ construction is motivated by Gaia observations of stellar populations, and its connection to the local dark-matter distribution depends on the relation between stellar and dark-matter debris. Streams and other asymmetric components can introduce a model-dependent characteristic-function phase and may invalidate the fixed $h$ proportionality. Precision forecasts should also include the escape-speed cutoff consistently throughout the model comparison, together with solar gravitational focusing~\cite{Lee2014} and more accurate orbital and terrestrial ephemerides.

Translating the accepted-event requirements into observing time or coupling sensitivity requires a detector-specific model. Finite interrogation windows weight velocities according to their detuning and can modify both the conditional coherence and the acceptance probability. Furthermore, the independent-event forecasts must be extended if successive interrogations retain correlations within the field coherence time. Environmental variations can produce solar-day or seasonal structure, so validation should combine off-resonance controls, phase-cycling tests, environmental monitoring, and checks that the predicted annual--sidereal modulation repeats across observing years.

Beyond model discrimination, measurements of both exchange quadratures and additional baseline lengths or orientations would constrain a larger set of velocity-space Fourier modes. Calibrated frequency-dependent responses provide complementary information through their velocity weighting. Such measurements can reduce parameter degeneracies and support regularized reconstruction, although a single rotating baseline does not determine a general three-dimensional velocity distribution. Multi-node networks could further expand the accessible modes and measurement protocols, with additional requirements for connectivity, calibration, and noise accounting. We leave those extensions to future work.

\vspace*{-0.2cm}
\section*{Acknowledgements}
\vspace*{-0.2cm}
We thank Hajime Fukuda for helpful discussion and correspondence. 
The work of MASA is supported in part by the US DOE under Award No DE-SC0024673.
The work of DK is supported in part by the NSF Grant No. PHY-2609913.
The work of KK is supported in part by the  NSF Grant No. PHY-2609759. 
The work of MP is supported by the National Research Foundation of Korea (NRF) grant funded by the Korea government (MSIT) (No.RS-2025-00564488). 
\vspace*{-0.2cm}

\appendix

\section{Escape-speed Cutoff}
\label{app:cutoff}

For an isotropic SHM truncated at $u<v_{\rm esc}$, angular integration gives
\begin{eqnarray}
 \Phi_{\rm R}^{\rm cut}(q)&=&
 \frac{4\pi}{N_{\rm R}\pi^{3/2}v_0^3}
 \int_0^{v_{\rm esc}}\dd u\,u^2e^{-u^2/v_0^2}j_0(qu), ~~\\
j_0(z)&=&\frac{\sin z}{z},
 \label{eq:cut-shm}
\end{eqnarray}
where $ N_{\rm R}=\operatorname{erf}(z)-\frac{2z}{\sqrt\pi}e^{-z^2}$ for $z=v_{\rm esc}/v_0$ ensures $\Phi_{\rm R}^{\rm cut}(0)=1$. The anisotropic counterpart is
\begin{equation}
 \Phi_{\rm S}^{\rm cut}(\vec  q)=\frac{1}{N_{\rm S}}
 \int_{u<v_{\rm esc}}\dd^3u\,
 \exp\!\left[-\frac{u_X^2}{2\sigma_r^2}
 -\frac{u_Y^2+u_Z^2}{2\sigma_t^2}+\ii\vec  q\cdot\vec  u\right],
 \label{eq:cut-sausage}
\end{equation}
where $N_{\rm S}$ is the corresponding numerical normalization, and $\sigma_r$ and $\sigma_t$ are the pre-cut Gaussian dispersions defined in Eq.~\eqref{eq:variances}.  The spherical boundary couples Cartesian components, so Eq.~\eqref{eq:cut-sausage} is evaluated numerically.  The finite-cutoff comparison in Fig.~\ref{fig:M_distributions} changes the benchmark response modestly and does not alter the modulation or degeneracy arguments.

\section{Earth-Fixed Baseline in Galactic Coordinates}
\label{app:geometry}

Let geographic latitude be $\lambda$ and define the local sidereal angle
\begin{equation}
 \theta_{\rm sid}=\frac{2\pi h}{24\,\mathrm{h}}.
 \label{eq:sidereal-angle}
\end{equation}
The outward unit vector from Earth's center to the site is 
\begin{equation}
 \hat{  u}=(\cos\lambda\cos\theta_{\rm sid},
 \cos\lambda\sin\theta_{\rm sid},\sin\lambda).
\end{equation}
Normalizing $\partial\hat{  u}/\partial\theta_{\rm sid}$ gives east, while $\partial\hat{  u}/\partial\lambda$ already has unit norm and gives north.  Thus the local east, north, and up basis in equatorial Cartesian coordinates is
\begin{align}
 \hat{  e}&=(-\sin\theta_{\rm sid},\cos\theta_{\rm sid},0),\\
 \hat{  n}&=(-\sin\lambda\cos\theta_{\rm sid},-\sin\lambda\sin\theta_{\rm sid},\cos\lambda),\\
 \hat{  u}&=(\cos\lambda\cos\theta_{\rm sid},\cos\lambda\sin\theta_{\rm sid},\sin\lambda).
 \label{eq:local-basis}
\end{align}
If azimuth $A_{\rm az}$ is measured east of north and altitude $a_{\rm alt}$ above the local horizon, the Earth-fixed directed baseline is
\begin{equation}
 \hat{r}_{\rm eq}=\cos a_{\rm alt}\sin A_{\rm az}\,\hat{  e}
 +\cos a_{\rm alt}\cos A_{\rm az}\,\hat{  n}
 +\sin a_{\rm alt}\,\hat{  u}.
 \label{eq:baseline-equatorial}
\end{equation}
We use the J2000 equatorial-to-Galactic rotation
\begin{widetext}
\begin{equation*}
 R_{\rm eq\to gal}=
 \begin{pmatrix}
 -0.0548755604&-0.8734370902&-0.4838350155\\
 +0.4941094279&-0.4448296300&+0.7469822445\\
 -0.8676661490&-0.1980763734&+0.4559837762
 \end{pmatrix},
 \qquad
 \hat{r}=R_{\rm eq\to gal}\hat{r}_{\rm eq}.
 \label{eq:eq-to-gal}
\end{equation*}
\end{widetext}
The displayed J2000 matrix~\cite{Liu2011} acts on equatorial column vectors; using its transpose would reverse the coordinate transformation.
The laboratory rotational velocity is
\begin{equation}
 \vec{v}_{\rm rot}(h)=\omega_\oplus R_\oplus\cos\lambda\,
 R_{\rm eq\to gal}\hat{  e}(h),
 \label{eq:rotation-velocity}
\end{equation}
with $\omega_\oplus=7.292115\times10^{-5}\ \mathrm{rad\,s^{-1}}$ and $R_\oplus=6378.137\ \mathrm{km}$.  The approximate orbital velocity used in the forecast is~\cite{Freese2013}
\begin{equation}
 \vec{v}_{\oplus}(d)=29.8\ \mathrm{km\,s^{-1}}
 \begin{pmatrix}
 0.9941\cos\varphi-0.0504\sin\varphi\\
 0.1088\cos\varphi+0.4946\sin\varphi\\
 0.0042\cos\varphi-0.8677\sin\varphi
 \end{pmatrix},
 \label{eq:orbital-velocity}
\end{equation}
where $ \varphi=0.0172(d-81)$ and $d$ is day of year.  A precision timestamp analysis should replace Eq.~\eqref{eq:orbital-velocity} with an ephemeris.

\section{Sensitivity Scaling and Parameter Degeneracies}
\label{sec:degeneracy}

We evaluate the full Bernoulli divergence in Eq.~\eqref{eq:kl} for the reported forecasts. The following small-difference expansion provides a useful explanation of the sensitivity scaling. For
$\delta\mu_k=\mu_k^{(H)}-\mu_k^{(T)}$ and
$|\mu_k^{(T)}|<1$,
\begin{equation}
 D_{\rm KL}(\mu_k^{(T)}\Vert\mu_k^{(H)})
 =\frac{(\delta\mu_k)^2}
 {2[1-(\mu_k^{(T)})^2]}
 +\mathcal O(\delta\mu_k^3).
 \label{eq:kl-quadratic}
\end{equation}
To illustrate this limit, we hold the anisotropy
parameter $\beta$, the offset $b$, and the effective
signal amplitude $g$ fixed. The mixture response and
the corresponding measured mean in bin $k$ are
\begin{align}
 \Mbar_{\eta,k}
 &=(1-\eta)\Mbar_{{\rm SHM},k}
   +\eta\Mbar_{S,k},\\
 \mu_{\eta,k}&=b+g\Mbar_{\eta,k},
\end{align}
where $\Mbar_{S,k}$ denotes the pure Sausage response.
Taking the mixture with fraction $\eta$ as the truth
and the SHM, corresponding to $\eta=0$, as the tested
model gives
\begin{equation}
 \delta\mu_k
 =\mu_{0,k}-\mu_{\eta,k}
 =-g\eta\frac{\partial\Mbar_{\eta,k}}{\partial\eta}.
\end{equation}
This identity holds separately in every observing bin.

For a fixed observing schedule, let
$w_k=N_k/N_{\rm post}$ denote the fraction of accepted
measurements assigned to bin $k$, with $\sum_k w_k=1$.
Substituting the quadratic expansion into
Eq.~\eqref{eq:asimov-info} gives
\begin{equation}
 q_{0,\mathrm{rep}}
 \simeq
 N_{\rm post}g^2\eta^2
 \sum_k w_k
 \frac{
 \left(\partial\Mbar_{\eta,k}/\partial\eta\right)^2
 }{1-\mu_{\eta,k}^2}.
\end{equation}
The $3\sigma$ criterion,
$q_{0,\mathrm{rep}}\simeq9$, therefore yields
\begin{equation}
 N_{3\sigma}^{\rm post}
 \simeq
 \frac{9}{g^2\eta^2}
 \left[
 \sum_k w_k
 \frac{
 \left(\partial\Mbar_{\eta,k}/\partial\eta\right)^2
 }{1-\mu_{\eta,k}^2}
 \right]^{-1}.
\end{equation}
For fixed $\beta$ and nonzero $g$, this expression gives
the approximate scaling
$N_{3\sigma}^{\rm post}\propto\eta^{-2}$ as $\eta\to0$,
provided the weighted sum approaches a finite,
nonzero value.

The connection to the parameter-precision criterion is
particularly transparent for a single bin. The sample
mean of $N_{\rm post}$ independent binary outcomes has
variance $(1-\mu_\eta^2)/N_{\rm post}$. Local error
propagation therefore gives
\begin{equation}
 \delta\eta_N
 \simeq
 \frac{\sqrt{1-\mu_\eta^2}}
 {\sqrt{N_{\rm post}}\,
  \left|g\,\partial\Mbar_\eta/\partial\eta\right|}.
\end{equation}
Consequently,
$q_{0,\mathrm{rep}}\simeq(\eta/\delta\eta_N)^2$.
In the ideal limit $b=0$ and $g=1$, the required
accepted-event count reduces to
\begin{equation}
 N_{3\sigma}^{\rm post}
 \simeq
 \frac{9(1-\Mbar^2)}
 {\eta^2\left|\partial\Mbar/\partial\eta\right|^2}.
\end{equation}
This has the same local sensitivity form as
Eq.~\eqref{eq:n3sigma-velocity}, with the variable of interest replaced by $\eta$.  

These expressions assume fixed nuisance parameters.
When $b$ or $g$ is profiled, the model difference must
first be minimized over their allowed values, as in
Eq.~\eqref{eq:asimov-info}. A single bin cannot
distinguish the models if an allowed shift of a free
offset absorbs the entire mean difference. Likewise,
fixed-LST annual observations retain the amplitude
degeneracy when $g$ is free and the required rescaling
is allowed. In either case, the profiled divergence
vanishes and no finite measurement count can distinguish
the models using those data alone.

\bibliography{refs.bib}

\end{document}